\documentclass[a4paper,reqno,12pt,final]{amsart}
\usepackage{amsmath,amsfonts,amssymb,amsthm,amstext,amscd}
\usepackage{eucal,array}
\usepackage{bbold}
\usepackage[latin1]{inputenc}
\usepackage[notref,notcite]{showkeys}
\DeclareMathOperator{\AdS}{AdS}

\DeclareMathOperator{\dvol}{dvol}
\DeclareMathOperator{\Ric}{Ric}
\DeclareMathOperator{\Weyl}{Weyl}
\DeclareMathOperator{\Gr}{Gr}
\DeclareMathOperator{\CW}{CW}
\DeclareMathOperator{\End}{End}
\renewcommand{\d}{\partial}
\newcommand{\RR}{\mathbb{R}}
\newcommand{\EE}{\mathbb{E}}
\newcommand{\SO}{\mathrm{SO}}

\newcommand{\ISO}{\mathrm{ISO}}
\newcommand{\fso}{\mathfrak{so}}
\newcommand{\fh}{\mathfrak{h}}

\newcommand{\eF}{\mathcal{F}}
\newcommand{\eC}{\mathcal{C}}
\newcommand{\eh}{\mathcal{H}}
\newcommand{\eE}{\mathcal{E}}
\newcommand{\eM}{\mathcal{M}}

\newcommand{\eA}{\mathcal{A}}
\newcommand{\half}{\tfrac{1}{2}}
\newcommand{\sto}{\rightsquigarrow}
\newcommand{\spb}{\$}
\calclayout
\numberwithin{equation}{section}
\begin{document}

\title{Penrose limits, supergravity and brane dynamics}
\author[Blau]{Matthias Blau}
\address{The Abdus Salam ICTP, Trieste, Italy}
\email{mblau@ictp.trieste.it}
\author[Figueroa-O'Farrill]{José Figueroa-O'Farrill}
\address{Department of Mathematics and Statistics, University of
Edinburgh, UK}
\email{j.m.figueroa@ed.ac.uk}
\author[Papadopoulos]{George Papadopoulos}
\address{Department of Mathematics, King's College, London, UK}
\email{gpapas@mth.kcl.ac.uk}
\begin{abstract}
  We investigate the Penrose limits of classical string and M-theory
  backgrounds.  We prove that the number of (super)symmetries of a
  supergravity background never decreases in the limit.  We classify
  all the possible Penrose limits of $\AdS \times S$ spacetimes and of
  supergravity brane solutions.  We also present the Penrose limits of
  various other solutions: intersecting branes, supersymmetric black
  holes and strings in diverse dimensions, and cosmological models.
  We explore the Penrose limit of an isometrically embedded spacetime
  and find a generalisation to spaces with more than one time.
  Finally, we show that the Penrose limit is a large tension limit for
  all branes including those with fields of Born--Infeld type.
\end{abstract}
\maketitle
\tableofcontents

\section{Introduction}

It has been shown recently that the maximally supersymmetric
backgrounds of M-theory and IIB superstring are related by a limiting
procedure, known as the \emph{Penrose limit}.  In particular, it has
been found that eleven-dimensional Minkowski spacetime and the
maximally supersymmetric Hpp-wave\footnote{The name \emph{Hpp-wave}
  was coined in \cite{FOPflux} to denote pp-waves with a homogeneous
  geometry, say $G/H$, and homogeneous (that is, $G$-invariant)
  fluxes.} of \cite{KG} (see also \cite{FOPflux}) can arise as Penrose
limits of the $\AdS_4\times S^7$ and $\AdS_7\times S^4$ solutions of
M-theory \cite{ShortLimits}.  In addition, ten-dimensional Minkowski
spacetime and the maximally supersymmetric Hpp-wave of \cite{NewIIB}
are Penrose limits of the $\AdS_5\times S^5$ solution of IIB
superstring theory. The other known maximally supersymmetric pp-wave
solutions in four \cite{KGd4} and five and six dimensions
\cite{Meessen} can also be obtained in this way.

As shown in \cite{MetsaevIIB}, the (Green--Schwarz) IIB superstring
on the maximally supersymmetric Hpp-wave solution is a free theory
after gauge-fixing, and can therefore be quantised exactly.  This fact
together with the above relation between the maximally supersymmetric
solutions of IIB supergravity were used in \cite{MaldaPL} to propose a
novel derivation of the spectrum of IIB superstrings on both Minkowski
and maximally supersymmetric Hpp-wave spacetimes from gauge theory
using the AdS/CFT correspondence.

In \cite{PenrosePlaneWave} Penrose showed that any spacetime (i.e.,
any solution of the Einstein field equations) has a limiting spacetime
which is a plane wave.  This limit can be thought of as a ``first
order approximation'' to the spacetime along a null geodesic.  The
limiting spacetime depends on the choice of null geodesic and hence a
spacetime can have more than one Penrose limit.  More recently, Güven
\cite{GuevenPlaneWave} extended Penrose's argument to show that any
solution of a supergravity theory has plane wave limits which are also
solutions.  This depends crucially on the local symmetries
(diffeomorphisms and gauge invariance) of supergravity theories as
well as the homogeneity of the supergravity action under a certain
constant rescaling of the fields, features already present in the
four-dimensional Einstein (or Einstein--Maxwell) action.

A novelty of the Penrose limit is that that string theory in pp-wave
backgrounds simplifies dramatically due to the existence of a natural
light-cone gauge, and in many cases can be quantised exactly.  String
theory on pp-wave NSR backgrounds has been considered before (see, for
example, \cite{TT} and references therein). In fact, the various
contractions of WZW models considered in
\cite{SfetsosPL,ORS,SfetsosTseytlinPL,TseytlinExact} are special cases
of a Penrose limit.

Since the Penrose limit relates in a precise way a general
supergravity background to a pp-wave, it allows us to understand
string theory on many backgrounds at least at a Penrose limit.  This
opens the possibility to probe string theory in backgrounds that were
hitherto inaccessible, for example backgrounds with RR fields, in the
background of various gravitational solitons, black holes and even
non-supersymmetric backgrounds.  The existence of different Penrose
limits is equally important, as every such limit is associated to a
perturbation theory with parameter the inverse of the string tension
\cite{ShortLimits} and different limits organise the perturbation
theory in different ways.  So it may be the case that what is a
perturbative effect in one perturbation expansion, may be viewed as
non-perturbative in another.

In this paper we shall investigate Penrose limits, following the
foundational work of Penrose and Güven, derive some of its basic
properties and explore the different Penrose limits of many
supergravity backgrounds.  We shall be primarily concerned with those
properties of a supergravity background which are inherited by its
Penrose limits.  By rephrasing the Penrose limit in the more general
context of ``limits of spacetimes'' introduced by Geroch
\cite{Geroch}, we shall show that the numbers both of symmetries and
supersymmetries never decrease in a Penrose limit.  We shall also show
that the symmetry superalgebra of a supergravity background gets
contracted in a Penrose limit.  Of course, the limiting background may
admit additional symmetries and supersymmetries not present in the
original background and in fact it can be shown that Penrose limits of
supergravity theories always preserve at least one half of the
supersymmetry.

After establishing the hereditary properties of the Penrose limit, we
classify the Penrose limits of $\AdS\times S$ spacetimes.  We will
show that such spacetimes can have only two distinct Penrose limits:
Minkowski spacetime or an Hpp-wave.  We shall then classify the
different Penrose limits of elementary brane solutions.  These
classifications follow by determining the orbits of the isometry group
of the spacetime on the space of null geodesics and invoking the
\emph{covariance property} of the Penrose limit, which says that null
geodesics which are related by an isometry yield Penrose limits which
are themselves isometric.

We will then exhibit Penrose limits of many supergravity solutions,
including
\begin{itemize}
\item M-branes, D-branes, NS-branes and their near-horizon limits;
\item intersecting brane solutions, in particular intersecting M-branes;
\item supersymmetric black hole solutions in four- and five-dimensions
  and string solutions in six dimensions as well as those of their
  near horizon geometries which are of $\AdS\times S$ type; and
\item cosmological models.
\end{itemize}

The latter case has been included to demonstrate the universality of
the Penrose limit and to point out a similarity between the Penrose
limit of the string solution and that of a cosmological model.  We
also investigate the Penrose limit of solutions that admit an
isometric embedding in flat space. This naturally leads to a
generalisation of the Penrose limit.  Finally we shall show that the
Penrose limit is a large tension limit for all brane probes in a
spacetime. This includes  branes with Born-Infeld type of worldvolume
fields like D-branes and the M5-brane. The precise dependence of the
D-brane probe actions on the string tension plays a key role.

\subsection{Contents and summary of main results}

This paper is organised as follows.  In Section~\ref{sec:PLimits} we
summarise some known facts about the Penrose limit in the context of
supergravity.  We define the Penrose limit, discuss the geometry of
limiting spacetime and relate it to pp-waves in both Rosen and
Brinkman coordinates.  We discuss the physical meaning of the limit
and derive the covariance property which will be crucial in later
sections.

In Section~\ref{sec:Lords} we introduce the concept of a
\emph{hereditary} property, modifying slightly the more general notion
introduced by Geroch \cite{Geroch}.  We discuss hereditary properties
involving the curvature tensor: showing that, for example, that
Einstein spaces have Ricci-flat limits, that conformal flatness is
hereditary and that so is the condition of being (locally) symmetric.
We show that the property of being a supergravity background is also
hereditary, so that solutions yield solutions in the limit, a fact
already observed by Güven.  We discuss isometries from two points of
view: one very explicit which shows why the symmetry algebra gets
contracted in the limit, and one more abstract which shows why its
dimension does not.  This is achieved by paraphrasing and slightly
generalising an argument originally due to Geroch.  This
generalisation consists in showing that the dimension of the space of
parallel sections of a family of vector bundles does not decrease in
the limit.  This is then applied to Killing vectors and to Killing
spinors to conclude, after a minor refinement, that the symmetry
(super)algebra of a Penrose limit is at least as large as that of the
original background.

In Section~\ref{sec:NHPL} we exhibit the maximally supersymmetric
Hpp-waves \cite{KG,FOPflux,NewIIB} of eleven-dimensional and IIB
supergravity as Penrose limits of the near-horizon geometries
\cite{GibbonsTownsend} of the D3 and M2/5 brane solutions. More
generally we show that all $\AdS\times S$ solutions have only two
different Penrose limits: Minkowski spacetime and the Hpp-waves.  In
addition we demonstrate explicitly that the isometries are hereditary
and in this way illustrate the contraction phenomenon.

Section~\ref{sec:branes} contains a classification of the possible
Penrose limits of elementary brane solutions of a supergravity theory.
Roughly speaking this is the ``cohomogeneity one'' analogue of the
classification in Section~\ref{sec:NHPL}.  As a result we do not
just obtain a finite number of non-isometric Penrose limits, but
rather a continuous family, labelled by the angular momentum of the
null geodesic along which we take the limit.  We illustrate these
results by explicit computations of the limits for a number of brane
solutions: D-branes, fundamental strings, NS-branes and M-branes.

Section~\ref{sec:intersections} is devoted to the Penrose limits of
intersecting brane solutions in ten and eleven dimensions.  After some
general remarks on the different Penrose limits of an intersecting
brane solution, we discuss explicitly the case of a triple pointlike
intersection of M2-branes and the intersection of two M2 and two M5
branes, along with their near-horizon geometries.

Section~\ref{sec:blackholes} considers the Penrose limits of
supersymmetric five- and four-dimensional black hole solutions arising
in toroidal compactification of string and M-theories.  We also
discuss the Penrose limit of a six-dimensional string solution.

Section~\ref{sec:frw} examines the Penrose limits of cosmological
models, focusing for definiteness on the four-dimensional
Friedmann--Robertson--Walker spacetime. For the spatially flat
cosmologies we find that the Penrose limit is a homogeneous Lorentzian
spacetime.

In Section~\ref{sec:embeddings} we discuss the Penrose limit of a
spacetime which is isometrically embedded in a flat pseudo-riemannian
space.  We will argue that there exists a generalisation of the notion
of Penrose limit for pseudoriemmanian spaces with signature $(s,t)$
where the limit is taken, not along null geodesics, but along totally
geodesic maximally isotropic submanifolds.  This is illustrated with
the near-horizon geometry of the four-dimensional Reissner--Nordström
black hole.  A similar discussion applies \emph{mutatis mutandis} to
the near-horizon geometry of the D3 brane.  We show, in passing, how
to isometrically embed the Hpp-wave metrics in flat space.

Finally in section~ \ref{sec:wd}, we show that Penrose limits are
large tension limits in the context of brane dynamics, including that
of branes with fields of Born-Infeld type. To establish this for the
case of D-branes, the precise dependence of D-brane actions on the
string tension is used.  We also find that every Penrose limit defines
a perturbation theory for the brane probes, including fundamental
strings.  Generically two different perturbation expansions are
related in a non-linear way. So a perturbative effect in one requires
the summation of an infinite number of graphs (terms) in the other.

\section{Penrose limit of supergravity theories}
\label{sec:PLimits}

In this section we review the Penrose limit of supergravity theories
along the lines of \cite{ShortLimits}.  We also review the physical
interpretation of the Penrose limit and we introduce a useful
covariance property.

\subsection{The Penrose--Güven limit}
\label{sec:PGLimit}

In this section we will review the Penrose limit as described by Güven
for backgrounds of supergravities in ten and eleven dimensions.

Let $(M,g)$ be a $D$-dimensional lorentzian spacetime.  According to
\cite{PenrosePlaneWave,GuevenPlaneWave} in the neighbourhood
of a segment of a null
geodesic $\gamma$ containing no conjugate points, it is possible to
introduce local coordinates $U,V,Y^i$ such that the metric takes the
form
\begin{equation}
  \label{eq:metric}
  g = dV \left(dU + \alpha dV + \sum_i \beta_i dY^i\right) +
  \sum_{i,j} C_{ij} dY^i dY^j~,
\end{equation}
where $\alpha$, $\beta_i$ and $C_{ij}$ are functions of all the
coordinates, and where $C_{ij}$ is a symmetric positive-definite
matrix.  The coordinate system breaks down as soon as $\det C = 0$,
signalling the existence of a conjugate point.  The coordinate $U$
is the affine parameter along a congruence of null geodesics labelled
by $V$ and $Y^i$.  The geodesic $\gamma$ is the one for which
$V=0=Y^i$.  Notice that the metric is characterised by the
conditions $g_{UV}=1/2$ and $g_{UU}=0=g_{UY^i}$ (we will also frequently
use coordinates in which $g_{UV}=1$). This means that the
vector field $\d/\d U$ is self-parallel and hence geodetic.  Also
its dual one-form $dV$ is closed so that the null geodesic congruence
into which $\gamma$ has been embedded is twist-free.

In ten- and eleven-dimensional supergravity theories there are other
fields besides the metric, such as the dilaton $\Phi$, gauge
potentials or more generally $p$-form potentials $A_p$ with
$(p+1)$-form field strengths.  The gauge potentials are defined up to
gauge transformations $A_p \mapsto A_p + d\Lambda_{p-1}$ in such a way
that the field strength $F_{p+1} = dA_p$ is gauge invariant.  It is
possible to use this gauge freedom in order to gauge away some of the
components of the $p$-form potentials.  Indeed, one can choose a gauge
in which
\begin{equation}
  \label{eq:gauge}
  i(\d/\d U) A = 0~,
\end{equation}
or in components
\begin{equation}
  A_{Ui_1i_2\dots i_{p-1}} = A_{UVi_1i_2\dots i_{p-2}} = 0~.
\end{equation}

The starting point of the Penrose limit is the data $(M,g,\Phi,A_p)$
defined in the neighbourhood of a conjugate-point-free segment of a null
geodesic $\gamma$ where $g$ and $A_p$ take the forms \eqref{eq:metric}
and \eqref{eq:gauge}, respectively.

We now introduce a positive real constant $\Omega >0$ and rescale the
coordinates as follows
\begin{equation}
  \label{eq:diffeo}
  U = u~, \qquad V = \Omega^2 v \qquad\text{and}\qquad Y^i =
\Omega y^i~.
\end{equation}
Substituting these expressions in the fields of the theory we obtain an
$\Omega$-dependent family of fields $g(\Omega)$, $\Phi(\Omega)$ and
$A_p(\Omega)$.  Let $\varphi_\Omega$ denote the
(local) diffeomorphism defined by \eqref{eq:diffeo}  and let us define
new fields
\begin{equation}
 \label{eq:fieldomega}
  g_\Omega = \Omega^{-2} \varphi_\Omega^* g \qquad
  \Phi_\Omega = \varphi_\Omega^* \Phi \qquad
  A_\Omega = \Omega^{-p} \varphi_\Omega^* A \implies
  F_\Omega = \Omega^{-p} \varphi_\Omega^* F~.
\end{equation}
These new fields $(g_\Omega, F_\Omega, \Phi_\Omega)$ are related to
the original fields $(g, F, \Phi)$ by a diffeomorphism and a
rescaling, and perhaps a gauge transformation.

Explicitly, $g_\Omega$ is
\begin{equation}
 \label{eq:gomega}
  \begin{split}
    g_{\Omega} =& dv du + \sum_{i,j} C_{ij}(u,\Omega y^i,\Omega^2 v)
    dy^i dy^j \\
    & + \Omega \sum_i \beta_i(u,\Omega y^i,\Omega^2 v) dy^i +
    \Omega^2\alpha(u,\Omega y^i,\Omega^2 v) (dv)^2~.
  \end{split}
\end{equation}

The coordinate and gauge choices \eqref{eq:metric} and
\eqref{eq:gauge} ensure that the following \emph{Penrose limit}
\cite{PenrosePlaneWave} (as extended by Güven \cite{GuevenPlaneWave}
to fields other than the metric) is well-defined:
\begin{equation}
  \label{eq:rescaling}
  \begin{aligned}[m]
    \bar g &= \lim_{\Omega\to 0} g_\Omega\\
    \bar \Phi &= \lim_{\Omega\to 0} \Phi_\Omega\\
    \bar A_p &= \lim_{\Omega\to 0} A_\Omega\\
    \bar F_{p+1} &= \lim_{\Omega\to 0} F_\Omega~.
  \end{aligned}
\end{equation}

By virtue of \eqref{eq:diffeo} the limiting fields only depend on the
coordinate $u$, which is the affine parameter along the null geodesic.
The resulting expression for the metric is
\begin{equation}
  \label{eq:PL}
  \bar g = du dv + \sum_{i,j} C_{ij}(u) dy^i dy^j~,
\end{equation}
where $C_{ij}(u)\equiv C_{ij}(u,0,0)$. We see that for the limit to
exist it is necessary that $g_{UU}=g_{UY^i}=0$. These conditions alone
are sufficient to ensure that $X=\d/\d U$ is self-parallel, $\nabla_X
X \propto X$, so that one may as well assume that $X$ is geodesic,
$\nabla_X X =0$. Taken together, these conditions then lead precisely
to the form of the metric given in \eqref{eq:metric}.

An obvious property of the metric \eqref{eq:PL} is that it is mapped
to itself under another Penrose limit along $\d/\d u$, and that the
Penrose limit of \eqref{eq:PL} along $\d/\d v$ is isometric to the
flat Minkowski metric.

The gauge potentials $\bar A_p$ only have components in the transverse
directions $y^i$:
\begin{equation}
  \label{eq:PLA}
  i(\d/\d u) \bar A_p = 0 =  i(\d/\d v) \bar A_p~,
\end{equation}
and the field strengths $\bar F_{p+1}$ are therefore of the form
\begin{equation}
  \label{eq:PLFS}
  \bar F_{p+1} = du \wedge \bar A_p(u)'~,
\end{equation}
where ${}'$ denotes $d/du$.  Notice in particular that $\bar F_{p+1}$
is always null.

So far we have not shown that a spacetime that emerges as the Penrose
limit of a solution of a supergravity theory is a solution of the same
theory. This is not apparent and it will be established later when the
hereditary properties of the Penrose limit are investigated. As we
shall see for the supergravity theories this follows because of the
invariance of the supergravity action under diffeomorphisms and gauge
transformations and the its homogeneity properties under the (overall)
scaling of the fields required for the Penrose limit and a continuity
argument.

The above expression for $\bar g$ is that of a pp-wave in Rosen
coordinates.  Generically it possess a ($2D{-}3$)-dimensional algebra of
isometries, isomorphic to a Heisenberg algebra, even though for
particular choices of $C_{ij}(u)$ the isometry algebra can be larger.
This Lie algebra is generated by the following Killing vectors:
\begin{equation}
  \label{eq:CWisomRosen}
  e_i = \frac{\d}{\d y^i}~, \quad e_+ = \frac{\d}{\d v}
  \quad\text{and}\quad e_i^* = y^i \frac{\d}{\d v} - \sum_j \int
  C^{ij}(u) du \frac{\d}{\d y^j}~,
\end{equation}
where $C^{ij}$ is the inverse of $C_{ij}$, which exists in the segment
of the null geodesic in which our coordinates are valid.  These
Killing vectors satisfy a Heisenberg Lie algebra
\begin{equation}
  \label{eq:heisenberg}
  \left[e_i,e_j\right]=0=\left[e_i^*,e_j^*\right] \qquad \left[e_i,
  e_j^*\right] = \delta_{ij} e_+
\end{equation}
with central element $e_+$.

\subsection{Penrose limits and pp-waves}
\label{sec:PLCW}

It is possible to change to Brinkman (also called harmonic)
coordinates in such a way that the Penrose limit metric \eqref{eq:PL}
metric takes the form
\begin{equation}
  \label{eq:CWtype}
  \bar g = 2 dx^+ dx^- + \left( \sum_{i,j} A_{ij}(x^-) x^i x^j \right)
   (dx^-)^2 + \sum_i dx^i dx^i~.
\end{equation}
When $A_{ij}$ is constant this metric describes a lorentzian symmetric
or Cahen--Wallach
space \cite{CahenWallach}.  Such spaces include the maximally
supersymmetric Hpp-waves of eleven-dimensional \cite{KG,FOPflux} and
IIB supergravity \cite{NewIIB}, namely
\begin{align}
  \label{eq:11dHpp}
  g_{11} &= 2 dx^+ dx^- - \left( \sum_{i,j=1}^3 \delta_{ij} x^i x^j +
    \tfrac14\sum_{i,j=4}^9 \delta_{ij} x^i x^j
  \right) (dx^-)^2 + \sum_{i=1}^9 dx^i dx^i\\
  \label{eq:IIBHpp}
  g_{\text{IIB}} &= 2 dx^+ dx^- - \left( \sum_{i,j=1}^8 \delta_{ij}
    x^i x^j \right) (dx^-)^2 + \sum_{i=1}^8 dx^i dx^i
\end{align}
(up to an overall scaling of $A_{ij}$ by a real positive constant
which can always be absorbed into a boost of $(x^+,x^-)$).

The explicit change of variables which takes the metric from Rosen to
Brinkman form is given by
\begin{equation}
  u = 2 x^- \qquad v = x^+ - \half \sum_{i,j} M_{ij}(x^-) x^i x^j \qquad
  y^i = \sum_{j} Q^i_j(x^-) x^j~,
\end{equation}
where $Q^i_j$ is an invertible matrix satisfying
\begin{equation}
  \label{eq:q}
  C_{ij}Q^i_k Q^j_l = \delta_{kl} \qquad\text{and}\qquad
  C_{ij} \left(Q'{}^i_j Q^j_l - Q'{}^i_k Q^j_l\right) = 0~,
\end{equation}
and
\begin{equation}
  M_{ij} = C_{kl}Q'{}^k_i Q^l_j~,
\end{equation}
which is symmetric by virtue of the second equation in \eqref{eq:q}.
Here a ${}'$ denotes differentiation with respect to $x^-$.
This equation guarantees that the limiting metric $\bar g$ has the
form \eqref{eq:CWtype}. The relation between $C_{ij}$ and $A_{ij}$ is
\begin{equation}
A_{ij} = -[C_{kl}Q'{}^l_j]'Q^k_i~.
\end{equation}

It remains to see that a $Q$ obeying \eqref{eq:q} always exists.
Since $C$ is symmetric it can be diagonalised at every point.
Moreover, since $C$ depends smoothly on the affine parameter $u$, we
can find a matrix $Q$ depending smoothly on $u$ such that the first
equation $Q C Q^t = I$ in \eqref{eq:q} is satisfied.  This does not
determine $Q$ uniquely, since we can always multiply on the left by an
orthogonal matrix $O$ depending smoothly on $u$.  Suppose that $Q_0$
has been chosen so that $Q_0 C Q_0^t = I$.  We claim it is possible to
find an orthogonal matrix $O$ such that $Q=O Q_0$ also satisfies the
second equation of \eqref{eq:q}.  This equation says that $M = Q' C Q$
is symmetric.  Let us then decompose $M = S + A$ into symmetric and
antisymmetric parts and similarly with $M_0 = Q_0' C Q_0 = S_0 + A_0$.
The equation for $O$ is then $A = O'O^t + O A_0 O^t = 0$, which is
equivalent to $O'=-A_0 O$.  This is a linear first-order differential
equation depending smoothly on $u$, and hence has a unique solution
for each initial value, at least for small enough $u$.

It is possible to rewrite the field strengths $\bar F_{p+1}$ given in
\eqref{eq:PLFS} in terms of Brinkman coordinates, to arrive at the
following expression:
\begin{equation}
  \label{eq:PLFSBrink}
  \bar F_{p+1} = \sum_{i_k,j_k} \frac{d}{dx^-} \bar A_{i_1i_2\dots
  i_p}(2x^-) Q^{i_1}_{j_1} Q^{i_2}_{j_2} \cdots Q^{i_p}_{j_p} dx^-
  \wedge dx^{j_1} \wedge dx^{j_2} \wedge \cdots \wedge dx^{j_p}~.
\end{equation}

A special case of this, frequently occurring in applications, is when
$C_{ij}$ is diagonal,
\begin{equation}
C_{ij}(u)= a_{i}^{2}(u)\delta_{ij}~.
\end{equation}
Then one can choose $Q^i_j=a_{i}(u)^{-1}\delta^i_j$ and finds
\begin{equation}
\label{eq:bi}
A_{ij}(x^-) = \frac{(a_i(x^-))''}{a_{i}(x^-)}\delta_{ij} \;.
\end{equation}
The (p+1)-form field strength becomes
\begin{equation}
\begin{aligned}[m]
  \bar F_{p+1} = \sum_{i_k} \frac{d}{dx^-} & \bar A_{i_1i_2\dots
  i_p}(2x^-) a_{i_1}(x^-)^{-1} a_{i_2}(x^-)^{-1} \cdots a_{i_p}(x^-)^{-1} \\
  &dx^-
  \wedge dx^{i_1} \wedge dx^{i_2} \wedge \cdots \wedge dx^{i_p}~.
  \end{aligned}
\end{equation}

In particular we learn from this that the Penrose limit of a metric
\eqref{eq:metric} with diagonal $C_{ij}$ is the flat metric if and only if
\begin{equation}
{a}_{i}(x^-)''=0~,
\end{equation}
i.e., if and only if
\begin{equation}
a_{i}(u) = b_{i} + c_{i}u~.
\end{equation}

We also learn that the Penrose limit is of Cahen--Wallach type
(constant nondegenerate $A_{ij}$) if and only if the functions
$a_{i}(u)$ are either trigonometric (for negative eigenvalues) or
hyperbolic (for positive eigenvalues). Indeed, if
\begin{equation}
a_{i}(u) = b_{i} \sin \mu_i u + c_{i} \cos \mu_i u
\end{equation}
or
\begin{equation}
a_{i}(u) = b_{i} \sinh \mu_i u + c_{i} \cosh \mu_i u~,
\end{equation}
then
\begin{equation}
a_{i}(u)'' = \mp\mu_i^2 a_{i}(u)~,
\end{equation}
and therefore $b_i(u)$ is constant,
\begin{equation}
b_i = \mp\mu_i^2~.
\end{equation}

\subsection{The physical effect of the Penrose limit}
\label{sec:PLblowup}

The physical interpretation of the Penrose limit is described by
Penrose as follows \cite{PenrosePlaneWave}:
\begin{quote}
  \emph{There is a `physical' interpretation of the above mathematical
    procedure, which is the following. We envisage a succession of
    observers travelling in the space-time $M$ whose world lines
    approach the null geodesic $\gamma$ more and more closely; so we
    picture these observers as travelling with greater and greater
    speeds, approaching that of light. As their speeds increase they
    must correspondingly recalibrate their clocks to run faster and
    faster (assuming that all space-time measurements are referred to
    clock measurements in the standard way), so that in the limit the
    clocks measure the affine parameter $x^0$ along $\gamma$. (Without
    clock recalibration a degenerate space-time metric would result.)
    In the limit the observers measure the space-time to have the
    plane wave structure $W_\gamma$.}
\end{quote}
In other words, the Penrose limit can be understood as a boost
followed by a commensurate uniform rescaling of the coordinates in
such a way that the affine parameter along the null geodesic remains
invariant.

A related physical effect of the Penrose limit is to blow up the
neighbourhood of a null geodesic.  Indeed, up to the coordinate
transformation \eqref{eq:diffeo}, the metric $g_{\Omega}$ differs from
the original spacetime metric $g$ only by the conformal rescaling
$g\to\Omega^{-2}g$.  Thus up to a coordinate transformation (which
becomes singular in the limit $\Omega\to 0$), the Penrose limit is a
large volume limit.

In particular, consider two points $P,Q$ is a neighbourhood of a null
geodesic and the (geodesic) distance
\begin{equation}
  d(P,Q; g) = \int^P_Q ds
\end{equation}
which is defined as the length of the shortest geodesic joining $P$
and $Q$, as measured with respect to the metric $g$.  Suppose now that
the geodesic joining $P$ and $Q$ is either timelike or spacelike, so
that $d(P,Q;g)\neq 0$.
The distance of the points $P,Q$ can be written as
\begin{equation}
  d(P,Q; g) = \Omega d(P,Q; \Omega^{-2} g(\Omega))~.
\end{equation}

Thus since $\Omega\ll 1$, we have that
\begin{equation}
  d(P,Q; g)=\Omega d(P,Q; \Omega^{-2} g(\Omega)) \ll d(P,Q; \Omega^{-2}
  g(\Omega))
\end{equation}
and therefore $d(P,Q; g) \ll d(P,Q; \bar g)$, where $d(P,Q; \bar g)$
is the distance of the two points as measured by the metric at the
Penrose limit.  Therefore, the distance of the two points $P,Q$ as
measured by the Penrose limit metric is thus much larger than that
measured with respect to the original spacetime metric $g$.  It is
worth mentioning though that if the points $P,Q$ are joined by a null
geodesic, they will remain separated by a null geodesic in the Penrose
limit.

\subsection{Covariance of the Penrose limit}
\label{sec:covariance}

Let $(M,g)$ be a lorentzian manifold.  A null geodesic $\gamma$ is
characterised (at least for small values of the affine parameter) by
specifying the initial position $\gamma(0)\in M$ and the initial
velocity $\Dot\gamma(0)\in T_{\gamma(0)}M$.  In fact, the Penrose
limit is only susceptible to the initial \emph{direction} of the
geodesic.  Indeed, if $\gamma_1$ and $\gamma_2$ are two null geodesics
starting at the same point but with collinear velocities; that is,
$\Dot\gamma_1(0) = \lambda \Dot\gamma_2(0)$ for some nonzero constant
$\lambda$, then the geodesics are related by a rescaling of the affine
parameter.  The resulting Penrose limits (see, e.g., \eqref{eq:PL})
are related by a rescaling of $u$, which can be reabsorbed in a
reciprocal rescaling of the conjugate coordinate $v$.  In other words,
the Penrose limit depends on the actual curved traced by the geodesic
and not on how it is parametrised.  We conclude that the Penrose limit
depends only on the data $(\gamma(0), [\Dot\gamma(0)])$, where
$\gamma(0)$ is a point in $M$ and $[\Dot\gamma(0)]$ is a point on the
(future-pointing, say) celestial sphere of $T_{\gamma(0)}M$, which is
the projectivisation of the nonzero future-pointing null vectors in
$T_{\gamma(0)}M$.

A fundamental property of the Penrose limit is that if two null
geodesics are related by an isometry, their Penrose limits are
themselves isometric.  We shall refer to this as the \emph{covariance
  property} of the Penrose limit.  This property is very useful in
determining the possible Penrose limits in spacetimes with a large
isometry group and will be used repeatedly in the bulk of the paper in
classifying (up to isometry) the possible Penrose limits of
supersymmetric backgrounds and their near-horizon geometries.

The covariance property holds because the isometry in question is by
assumption $\Omega$-independent and will therefore continue to exist
when $\Omega=0$.  By contrast, if two metrics $g_{\Omega}$ and
$h_{\Omega}$ are related by an $\Omega$-dependent isometry, then their
Penrose limits need not be isometric because the isometry between them
could become singular in the limit. For example, the metrics
$g_{\Omega}$ and $\Omega^{-2}g$ are isometric for all finite values of
$\Omega$, being related by the $\Omega$-dependent coordinate
transformation \eqref{eq:diffeo}. In the limit $\Omega\to 0$, however,
the limit of $g_{\Omega}$ is the Penrose limit, which is generically
not isometric to the naive large volume limit, i.e., the limit as
$\Omega\to 0$ of $\Omega^{-2}g$ (which is well defined if one combines
it with the coordinate transformation $(U,V,Y^i)=\Omega(u,v,y^i)$).

We see that the notion of a Penrose limit or, more generally,
that of a limit of a family of spacetimes in the sense discussed
in \cite{Geroch} (see also Section~\ref{sec:setup}),
is not invariant under coordinate transformations
depending on the parameter labelling the family of spacetimes. This
accounts for the fact that the definition of the Penrose limit given
in Section~\ref{sec:PGLimit} looks rather non-covariant. Indeed it is,
and it necessarily has to be. The situation would be different if we were
interested in all possible limits of a family of spacetimes \cite{Geroch},
but we are not.

\section{Hereditary properties of Penrose limits}
\label{sec:Lords}

In this section we discuss those properties of a supergravity
background that are inherited by all its Penrose limits.

\subsection{The set-up}
\label{sec:setup}

As mentioned in the introduction, Penrose limits of solutions of
supergravity theories are also solutions by virtue of the homogeneity
of the action under scaling. This observation is due to Penrose
\cite{PenrosePlaneWave} originally and to Güven \cite{GuevenPlaneWave}
in the supergravity context. We will re-establish this below.  It is
then of interest to investigate whether the Penrose limit of a
solution inherits some other properties (e.g., isometries,
supersymmetry) of the original solution.

The appropriate framework for addressing these questions has been
introduced by Geroch in 1969 \cite{Geroch}. In this somewhat more
general context one considers a (one-parameter) family of spacetimes
$(M_\Omega,g_\Omega)$ for $\Omega >0$ and tries to make sense and study
the properties of the limit spacetime as $\Omega\to 0$.

The reason for considering such a one-parameter family instead of a
one-parameter family of metrics $g_\Omega$ on a fixed spacetime $M$
is that, as we saw in Section~\ref{sec:covariance}, the limit is
not invariant under $\Omega$-dependent coordinate transformations.
Hence, if one is interested in studying \emph{all} limits of a given
one-parameter family of space-times, one should not presuppose an a priori
identification of points in the different $M_\Omega$. This evidently
leads to some technical complications in defining \emph{the} limit of
a family of spacetimes which are analysed and resolved in \cite{Geroch}.

In our case, however, we can sidestep these technical problems and work
with a fixed spacetime $M$ because we are not interested in studying
all possible limits but we are singling out a particular limit,
namely the Penrose limit. Evidently, the coordinate transformation
\eqref{eq:diffeo} provides us precisely with an identification of points
between $M_{\Omega=1}=M$ and $M_\Omega$, and then the family of spacetimes
defined by \eqref{eq:gomega} refers to a fixed spacetime $M$ equipped with a
one-parameter family of metrics $g_\Omega$.

In either case it is convenient to  consider the family
$(M_\Omega,g_\Omega)$ of $D$-dimensional spacetimes with $\Omega>0$
as a $(D+1)$-dimensional manifold $\eM$ equipped with a degenerate
metric $g_\Omega$ and a scalar field, namely $\Omega$, whose level surfaces
are the $M_\Omega$. Questions about the limit of $(M_\Omega,g_\Omega)$ are
then questions about the boundary $\d\eM$ of $\eM$.

Geroch calls a property of spacetimes \emph{hereditary} if, whenever a
family of spacetimes have that property, all the limits of this family
also have this property. For present purposes we find it convenient to
slightly modify this definition. We will call a property of a
supergravity configuration \emph{hereditary} if, whenever a
supergravity configuration has this property, all the Penrose limits
of this configuration also have this property.

This definition differs in three respects from Geroch's definition:
\begin{enumerate}
\item First of all, instead of referring just to a property of the
  spacetime $(M,g)$, the topological, causal and metric properties of
  the spacetime manifold, we consider all the supergravity fields and
  talk about the properties of a supergravity configuration
  $(M,g,\Psi)$ where $\Psi$ collectively refers to the supergravity
  matter fields.
\item Moreover, instead of talking about all limits, we only consider
  Penrose limits.
\item Finally, instead of talking about a property of a family of
  spacetimes, we only refer to a property of the initial supergravity
  configuration.
\end{enumerate}
The reason for the first two modifications is obvious. The reason for
the third modification is that we are usually not given a family of
spacetimes but one particular supergravity configuration, and we would
then like to study the properties of this configuration in all
possible Penrose limits, i.e., for all null geodesics.

This means that to check if a certain property of a supergravity
configuration $(M,g,\Psi)$ is hereditary, we first have to check if it
holds for $(M_\Omega=M,g_\Omega,\Psi_\Omega)$ for $\Omega >0$.  Now
$(g_\Omega,\Psi_\Omega)$ differ from $(g,\Psi)$ by a scaling and a
diffeomorphism (and possibly a gauge transformation), and we are only
interested in generally covariant and gauge invariant properties of
$(M,g,\Psi)$.  Thus in practice this amounts to checking if the
property of interest is invariant under a finite scaling of the fields
$(g,\Psi)$ before investigating what happens as $\Omega \to 0$.

It is clear that any property which is hereditary in the sense of
Geroch (once that definition is extended in a straightforward way to
include matter fields) is also hereditary in our sense.  However, as
Penrose limits are a rather special class of limits of spacetimes, it is
conceivable that there are hereditary properties of Penrose limits which
are not hereditary in general. Even though these are likely to exist,
all the hereditary properties we will discuss in this section are also
hereditary in the sense of Geroch.

Certain spacetime (or supergravity) properties are obviously
hereditary, for example those that can be expressed in terms of
tensorial equations for the Riemann tensor---see
Section~\ref{sec:lordriem} below.  Other hereditary properties are
less obvious. For example, when it comes to isometries, one could
imagine that in the (Penrose) limit of family of spacetimes, all
possessing a certain number $n$ of Killing vectors, one finds less
linearly independent Killing vectors simply because some Killing
vectors which happen to be linearly independent for all $\Omega > 0$
cease to be linearly independent at $\Omega = 0$.  This is at least
what a direct approach to the problem would suggest as being possible.

However, a very elegant and powerful argument due to Geroch \cite{Geroch},
which we will recall and generalise slightly below, establishes that
the number of linearly independent Killing vectors can never decrease
in the limit.  This argument has the additional virtue of being readily
applicable to Killing spinors and supersymmetries. As a consequence we
will also establish that the number of supersymmetries preserved by
a supergravity configuration can never decrease in the Penrose limit.
It is of course possible, and it is often the case, that the Penrose
limit actually has more (super-)symmetries than the original
spacetime.  Indeed, as we will see later, the Penrose limit preserves
at least one half of the supersymmetry.

\subsection{Hereditary properties involving the curvature tensor}
\label{sec:lordriem}

One of the most elementary and basic hereditary properties of any
family $(M_{\Omega},g_{\Omega})$ of space-times is the following
\cite{Geroch}: If there is some tensor field constructed from the
Riemann tensor and its derivatives which vanishes for all $\Omega >0$,
then it also vanishes for $\Omega =0$.

In particular, the Penrose limit\footnote{Of course there is no such
thing as \emph{the} Penrose limit, as it in general depends on the
choice of null geodesic $\gamma$. When talking about \emph{the Penrose
limit} without specifying a particular $\gamma$, we mean \emph{any
Penrose limit}.} of a Ricci-flat space-time is Ricci-flat, and the
Penrose limit of a conformally flat space-time (vanishing Weyl tensor,
$\Weyl(g_{\Omega})=0$) is conformally flat.

However, the Penrose limit of an Einstein manifold with fixed non-zero
cosmological constant or scalar curvature is not of the same type (as
the Ricci scalar, unlike the Ricci tensor, is not scale-invariant).
Rather, if $g$ is Einstein, so that $\Ric(g) = \Lambda g$, then
$\Ric(g_\Omega) = \Lambda \Omega^2 g_\Omega$, and in the limit
$\Ric(\bar g) = 0$.  In other words, the Penrose limit of an Einstein
space is always Ricci-flat.  Conversely, if every Penrose limit of a
space-time is Ricci-flat, then the original space-time is Einstein
\cite{PenrosePlaneWave}.

Similarly the Penrose limit of a locally symmetric space is locally
symmetric.  Indeed, the vanishing of the covariant derivative of the
Riemann curvature tensor is a hereditary property.  The only metrics
of type \eqref{eq:CWtype} which are locally symmetric are the
Cahen--Wallach metrics where $A_{ij}$ is constant.  It follows that
the Penrose limit of an $\AdS\times S$ space, which is a symmetric
space, is a Cahen--Wallach spacetime but with a possibly degenerate
$A_{ij}$; that is, it is locally isometric to a product of an
indecomposable Cahen--Wallach spacetime with a flat space.  Penrose
limits of $\AdS \times S$ are discussed in Section~\ref{sec:NHPL},
which contains a more precise statement.

\subsection{Supergravity equations of motion}
\label{sec:sugraeom}

We saw in Section~\ref{sec:PLimits} that two of the ingredients of the
Penrose limit are a change of coordinates \eqref{eq:diffeo} and a
rescaling \eqref{eq:rescaling} of the fields.  The change of variables
is a diffeomorphism for arbitrary nonzero $\Omega$.  The limit
consists in taking the limit $\Omega \to 0$, where the change of
coordinates becomes singular, but having rescaled the fields in such a
way that the limit exists.

What makes this limit interesting is the
fact that the supergravity actions considered here are homogeneous
under the rescaling.  Indeed, let the metric rescale as
\begin{equation}
  g\sto\Omega^{-2} g
\end{equation}
and the $p$-form gauge potentials as
\begin{equation}
  A_p\sto\Omega^{-p} A_p~.
\end{equation}
This implies that the field strengths
rescale as
\begin{equation}
  F_{p+1} \sto \Omega^{-p} F_{p+1}
\end{equation}
and also that the Hodge star $\star$ acting on $p$-forms in $D$ dimensions scales as
\begin{equation}
  \star_p\sto\Omega^{2p-D}\star_p~.
\end{equation}
In particular, the Hodge star
operator in an even-dimensional space is conformally invariant when
acting on middle-dimensional forms.  This means that self-duality
conditions which do not follow from an action principle, like that of
the $5$-form field strength in IIB supergravity, for instance, are
invariant under rescaling.

Under the rescaling of the metric the $D$-dimensional
Einstein--Hilbert action is homogeneous with degree $2-D$:
\begin{equation}
  \int d^Dx \sqrt{|\det g|} R \sto \Omega^{2-D} \int d^Dx \sqrt{|\det
  g|} R~.
\end{equation}
Similarly the Maxwell action is homogeneous of the same degree:
\begin{equation}
  \int F_{p+1} \wedge \star F_{p+1} \sto \Omega^{2-D}   \int F_{p+1}
  \wedge \star F_{p+1}~,
\end{equation}
and the same is true of the dilaton action, which can be thought of as
the $p=0$ case of the above action:
\begin{equation}
  \int d\Phi \wedge \star d\Phi \sto \Omega^{2-D} \int d\Phi \wedge
  \star d\Phi~.
\end{equation}
As the dilaton scales trivially, the above results remain true when
there are non-minimal couplings of the dilaton to the IIB metric (in
the string frame) and the RR-fields.

Likewise, on dimensional grounds, the cubic Chern-Simons terms of
supergravity in $D$ dimensions, like those of eleven-dimensional
supergravity (for $p=3$) and of five-dimensional $N{=}2$ supergravity
(for $p=1$),
\begin{equation}
  \int F_{p+1} \wedge F_{p+1} \wedge A_p~,
\end{equation}
and generically of the form
\begin{equation}
  \int  dA_p^{(1)}\wedge dA_q^{(2)}\wedge A_{D{-}2{-}p{-}q}^{(3)}~,
\end{equation}
are also homogeneous of the same degree,
\begin{equation}
  \int  dA_p^{(1)}\wedge dA_q^{(2)}\wedge A_{D{-}2{-}p{-}q}^{(3)}
\sto \Omega^{2-D}
  \int  dA_p^{(1)}\wedge dA_q^{(2)}\wedge A_{D{-}2{-}p{-}q}^{(3)}~.
\end{equation}

The fermionic terms in the supergravity action also scale
homogeneously with weight $2-D$ provided that the fermionic fields
scale appropriately.  Consider the kinetic term of a gravitino:
\begin{equation}
  \int d^Dx \sqrt{|\det g|} \bar\psi_M \Gamma^{MNP} \nabla_N \psi_P~.
\end{equation}
This term scales with weight $2-D$ provided that the gravitino scales
with $-1/2$.  Similarly, the kinetic term of other fermions
\begin{equation}
  \int d^Dx \sqrt{|\det g|} \bar\lambda \Gamma^{M} \nabla_M \lambda
\end{equation}
scales with weight $2-D$ if and only if the fermions scale with weight
$1/2$.  With these conventions, other fermionic terms in which
fermions couple non-minimally to the field strengths $F_{p+1}$ again
scale homogeneously with weight $2-D$.

These properties imply that the field equations are homogeneous.  In
particular, if a bosonic configuration $(g,F,\Phi)$ satisfies the
equations of motion so will the rescaled fields.

The new fields $(g_\Omega, F_\Omega, \Phi_\Omega)$ defined in
\eqref{eq:fieldomega} are related to the original fields $(g, F,
\Phi)$ not just by a rescaling but also by a diffeomorphism and
perhaps a gauge transformation to bring the fields into a form in
which the limit $\Omega \to 0$ exists.  But since the equations of
motion are covariant under diffeomorphisms and gauge transformations
as well as under the rescaling, the new fields $(g_\Omega, F_\Omega,
\Phi_\Omega)$ will satisfy the equations of motion for any $\Omega >
0$ if the original fields $(g, F, \Phi)$ do.  A general continuity
argument now guarantees that the limiting fields
\begin{equation}
  \bar g = \lim_{\Omega \to 0} g_\Omega \qquad \bar \Phi = \lim_{\Omega
  \to 0}\Phi_\Omega \qquad \bar F = \lim_{\Omega \to 0} F_\Omega
\end{equation}
also satisfy the equations of motion\footnote{To make this  and the
various other continuity arguments below rigorous, one has to put an
appropriate topological structure on the space of the various objects,
like the space of solutions, that are involved in arguments and show that
the limiting processes described are continuous.  However we shall not
attempt this here.}. Thus the Penrose limit of a supergravity solution
is a (possibly new) solution of the supergravity equations of motion.

\subsection{Isometries I: preliminary considerations}
\label{sec:isometries1}

We now come to the more subtle issue of isometries (and supersymmetries).
To set the stage, let $\xi$ be a Killing vector of the metric $g$. Then
in the rescaled coordinates \eqref{eq:diffeo}, $\xi$ acquires a dependence
on $\Omega$, $\xi\to\xi(\Omega)$ and $\xi(\Omega)$ is a Killing vector
for the transformed metric $g(\Omega)$ as well as for the transformed
and rescaled metric $\Omega^{-2}g(\Omega)$.
Hence in the limit
\begin{equation}
\bar\xi=\lim_{\Omega\to 0} \Omega^{\Delta_\xi} \xi(\Omega)
\end{equation}
is a non-trivial
Killing vector of the limiting metric $\bar g$ provided that $\Delta_{\xi}\in\RR$
can be chosen so that the above limit exists, i.e., such that the limit is both
non-singular and non-zero.

We will now show that such a $\Delta_{\xi}$ always exists.
If the Killing vector $\xi$ in the local
coordinates adapted to a null geodesic is
\begin{equation}
\xi = \alpha(U,V,Y^i)\d_U + \beta(U,V,Y^i)\d_V + \gamma^i(U,V,Y^i)\d_{Y^i}~,
\end{equation}
then $\xi(\Omega)$ is
\begin{equation}
\xi(\Omega) = \alpha(u,\Omega^2v,\Omega y^i)\d_u +\Omega^{-2}
 \beta(u,\Omega^2v,\Omega y^i)\d_v +
\Omega^{-1}\gamma^i(u,\Omega^2v,\Omega y^i)\d_{y^i}~.
\end{equation}
For sufficiently small $\Omega$ we can expand the Killing vector about
$\Omega=0$, i.e., about the geodesic $(u,v=0,y^i=0)$, and find
\begin{equation}
\Omega^2\xi(\Omega) = \bar\beta(u)\d_v +\Omega(\bar\gamma^i(u)\d_{y^i}
+ y^{i}\d_{y^i}\bar\beta(u)\d_v) +  \ldots,
\end{equation}
where $\bar\beta(u) = \beta(u,0,0)$ etc. Now let $\Omega^{k_\xi}$ be
the coefficient of the first non-zero term on the right hand side of
this Taylor expansion. Thus $k_{\xi}\geq 0$ and
\begin{equation}
\lim_{\Omega\to 0}\Omega^{2-k_\xi}\xi(\Omega)
\end{equation}
is finite and non-zero. This shows that $\bar\xi$ is finite and
non-zero with the choice $\Delta_\xi = 2-k_\xi\leq 2$.

The infinitesimal symmetries of a supergravity
background $(M,g,\Psi)$ are given by Killing vectors which in addition
leave invariant the other fields in the background, i.e.\ we require
\begin{equation}
  L_\xi \Psi = 0~.
\end{equation}
Without loss of generality, we can assume that each $\Psi$ is in the Penrose-G\"uven
gauge. Then $\Psi$ has a well defined Penrose-G\"uven limit
\begin{equation}
\bar\Psi = \lim_{\Omega\to 0}\Psi_{\Omega}=
\lim_{\Omega\to 0}\Omega^{\Delta_{\Psi}}\Psi
\end{equation}
(which may or may not be non-zero). To show that this symmetry of the background
is preserved in the Penrose limit, we observe that
\begin{equation}
  L_{\bar\xi}\bar\Psi =
\lim_{\Omega\to 0} \Omega^{\Delta_\xi +\Delta_\Psi}L_\xi\Psi =0~.
\end{equation}

Now consider two linearly independent Killing vectors $\xi_1$ and
$\xi_2$ of $(M,g)$. It is of course perfectly possible that
$\xi_1(\Omega)$ and $\xi_2(\Omega)$ are linearly independent for
$\Omega> 0$ but that their leading order terms in a small-$\Omega$
expansion are linearly dependent (see e.g., the example in
Section~\ref{sec:isometriestoo}).  Let us assume that this happens
and, without loss of generality, that the leading order terms are in
fact equal. Then $\bar\xi_1=\bar\xi_2$ and we appear to have lost a
Killing vector upon taking the limit. However, let us now consider the
difference
\begin{equation}
  \xi_-(\Omega)=\xi_1(\Omega) -\xi_2(\Omega)~.
\end{equation}
In this difference the leading order term proportional to
$\bar\xi_1-\bar\xi_2=0$ drops out and the first non-vanishing term,
with
\begin{equation}
  \Delta_{\xi_-} < \Delta_{\xi_1}=\Delta_{\xi_2}~,
\end{equation}
defines a Killing vector $\bar\xi_-$. If this Killing vector is
linearly independent of $\bar\xi_1$, then the procedure stops here.
If not, one needs to iterate this procedure. It is possible to show
that in this way one eventually ends up with two linearly independent
Killing vectors of the Penrose limit spacetime $(M,\bar g)$.

However, this procedure is not very enlightening, and the result is
actually a special case of a general result by Geroch which states
that the number of linearly independent Killing vectors can never
decrease in the limit of a family of spacetimes possessing a fixed
number of linearly independent Killing vectors.  Geroch's elegant
argument, which we will recall below, also generalises in a
straightforward way to Killing spinors and supersymmetries.

Anticipating this result, let us make two more remarks. The first is
that because different Killing vectors may have to be rescaled with
different values of $\Delta$, the isometry algebra may get contracted
in the limit.  We will see an example of this phenomenon in
Section~\ref{sec:isometriestoo} when discussing the isometry algebra
of Penrose limits of spacetimes of the form $\AdS\times S$.

Let us also note that, even if the original metric has no isometries,
the Penrose limit always does. This is because, as we saw in
Section~\ref{sec:PLimits}, $D$-dimensional metrics of the form
\eqref{eq:PL} always possess a ($2D{-}3$)-dimensional algebra of
isometries, isomorphic to a Heisenberg algebra. Since generically
these isometries have no counterpart in the original space-time and
only arise at $\Omega = 0$, one should then not think of these
isometries as hereditary but rather as a \emph{post mortem} effect.

Thus, starting with a spacetime $(M,g)$ with $n$ linearly independent
Killing vectors, the number of linearly independent Killing vectors of
the Penrose limit space-time is always at least as large as $\max
(n,2D{-}3)$.

\subsection{Isometries II: Geroch's argument}
\label{sec:isometries2}

We will now study the fate of isometries in the more general context
of hereditary properties of limits of spacetimes.  We begin with a
generalisation of an argument due to Geroch for the hereditary
property of isometries.  This will allow us to easily extend this
argument from isometries to supersymmetries.  The crucial observation
is that (super)symmetries of a supergravity background are in
one-to-one correspondence with Killing vectors and Killing spinors
subject, perhaps, to algebraic conditions; and that the condition of a
vector or a spinor being Killing can be rephrased in terms of a
section of a certain vector bundle being parallel: a subbundle of the
tensor bundle in the case of Killing vectors and the spinor bundle in the
case of Killing spinors.  We therefore start by discussing what
happens to parallel sections of vector bundles under limits such as
the one discussed in Section~\ref{sec:setup}.

\subsubsection{A slight generalisation of Geroch's argument}
\label{sec:geroch}

We will be concerned here with vector bundles with connection defined
on $\eM$ (see Section~\ref{sec:setup}) which moreover extend to the
boundary $M_0$.  More precisely, let $E_\Omega \to M_\Omega$, $\Omega
> 0$ be a smooth family of rank $k$ vector bundles with connection
$D^\Omega$.  We will assume that the limit $\Omega \to 0$ exists, so
that $E_0$ is a smooth rank-$k$ vector bundle on $M_0$ with connection
$D^0$.  This hypothesis will be justified for each of the cases to
which we will apply the results of this section.  Indeed, the bundles
(and their connections) under consideration will only depend on the
metric and the other bosonic fields of the supergravity theory, which
have well-defined Penrose limits.

The connection $D^\Omega$ defines a notion of parallel transport along
curves $c_\Omega : I \to M_\Omega$, and by restricting to closed
curves, a notion of holonomy.  Parallel sections of $E_\Omega$, if
they exist, define a rank-$k'$ subbundle $E'_\Omega \subset E_\Omega$:
the fibre $E'_\Omega(p_\Omega)$ at $p_\Omega \in M_\Omega$ is the
subspace of $E_\Omega(p_\Omega)$ spanned by the values at $p_\Omega$
of the parallel sections, equivalently of those sections which are
invariant under the holonomy group of $D^\Omega$ at $p_\Omega$.  We
would like to investigate the limit as $\Omega\to 0$ of the family
$(E_\Omega,D^\Omega)$ and in particular to say something about the
rank of $E'_0 \subset E_0$.  We will see, in fact, that the rank of
$E'_0$ is not smaller than the rank of $E'_\Omega$.

To see this, let us choose a point $p_0 \in M_0$ and a path $t \mapsto
p_t$ in $\eM$, such that for each $\Omega>0$, $p_\Omega \in M_\Omega$
and such that the limit $\lim_{\Omega \to 0} p_\Omega = p_0$, as the
notation suggests.  We can always trivialise the bundle along the
path, and in this way identify the fibres $E_\Omega(p_\Omega)$ with a
fixed $k$-dimensional vector space $E$.  The fibres
$E'_\Omega(p_\Omega)$ define a family of $k'$-dimensional subspaces of
$E$, and hence a path in the Grassmannian $\Gr(k',E)$ of
$k'$-dimensional planes in $E$.  Because the Grassmannian is compact,
this path has a limit point in the Grassmannian as $\Omega \to 0$, and
thus we obtain a $k'$-dimensional subspace $E'_0(p_0)$ of $E_0(p_0)$.
Doing this for all $p_0$ in the boundary $M_0$ we obtain a rank-$k'$
subbundle $E'_0 \subset E_0$.

We will now show that $E'_0$ is left invariant by the holonomy group
of the connection $D^0$.  To this end let $c_0$ be a closed curve
through $p_0$ in $M_0$ and let $c_\Omega$ be a family of closed curves
through $p_\Omega$ in $M_\Omega$ in such a way that as $\Omega \to 0$,
$p_\Omega \to p_0$ and $c_\Omega \to c_0$.  Now consider a basis for
$E'_\Omega$ near $p_\Omega$ made out of parallel sections of
$E_\Omega$ and yielding a basis for $E'_0$ in the limit $\Omega\to 0$.
(The existence of such a basis is guaranteed by the argument in the
previous paragraph.)  Their parallel transport around $c_\Omega$ is
trivial for all $\Omega>0$, hence by continuity the parallel transport
along $c_0$ of a basis for $E'_0$ is again trivial.  This shows that
all elements of $E'(p_0)$ can be integrated to parallel sections of
$E_0 \to M_0$.

In practice we will be interested in parallel sections satisfying
additional linear equations.  For example, the infinitesimal
symmetries of a supergravity background are Killing vectors which, in
addition to the metric, also leave invariant the other fields in the
background.  These conditions single out a linear subspace of the
parallel sections and we will be interested in the fate of this
subspace in the limit.  This requires a slight refinement of the above
argument, which we now detail.

Let $C_\Omega$ be a a family of linear conditions on sections of
$E_\Omega \to M_\Omega$, depending smoothly on $\Omega$ and having a
well-defined limit as $\Omega \to 0$.  In practice, $C_\Omega$ will
depend on the supergravity fields and hence the limit $\Omega\to 0$ is
well-defined by virtue of these fields having a well-defined
Penrose--Güven limit.  For every point $p_\Omega$ let
$E''_\Omega(p_\Omega) \subset E_\Omega(p_\Omega)$ denote the linear
subspace spanned by the values $\psi_\Omega(p_\Omega)$ at $p_\Omega$
of parallel sections $\psi_\Omega$ of $E_\Omega$ which in addition
satisfy the condition
\begin{equation}
  \label{eq:extraconditions}
  C_\Omega \psi_\Omega = 0~.
\end{equation}
These subspaces define a rank-$k''$ sub-bundle $E''_\Omega \to
M_\Omega$.  Repeating the argument above for this subbundle we find
that for every point $p_0 \in M_0$ in the boundary of $\eM$, we obtain
a $k''$-dimensional subspace $E''(p_0)$ of $E(p_0)$ spanned by the
values at $p_0$ of parallel sections of $E_0 \to M_0$.  Now suppose
that $\{\psi^{(1)}_\Omega,\dots,\psi^{(k'')}_\Omega\}$ is a frame for
$E''_\Omega \to M_\Omega$ yielding in the limit $\Omega\to 0$ a frame
for $E''_0$.  Since each $\psi^{(i)}_\Omega$ obeys equation
\eqref{eq:extraconditions}, continuity implies that in the limit $C_0
\psi_0^{(i)} = 0$.  In summary, the dimension of the space of
$D^\Omega$-parallel sections $\psi_\Omega$ of $E_\Omega\to M_\Omega$
obeying the conditions \eqref{eq:extraconditions} cannot decrease in
the limit $\Omega \to 0$.

We will apply this argument both to isometries and supersymmetries, by
realising them as parallel sections of appropriate vector bundles with
connection, perhaps subject to additional linear conditions.

\subsubsection{Killing transport}
\label{sec:killingtrans}

We now describe a useful local characterisation of Killing vectors as
parallel sections of a vector bundle with connection (see, for
example, \cite{KostantHol,Geroch}).

Let $X$ be any vector field on a connected $n$-dimensional spacetime
$(M,g)$ and let $A_X$ denote the map taking a vector field $Y$ to
$\nabla_Y X$.  This map is tensorial (that is, $C^\infty(M)$-linear)
and hence defines a section of $\End(TM) \cong T^*M \otimes TM$.  A
vector field $\xi$ is a Killing vector if and only if $A_\xi$ is
skew-symmetric relative to $g$; indeed, a vector $\xi$ is Killing if
and only if
\begin{equation}
  g(\nabla_X\xi, Y) = - g(X, \nabla_Y \xi)
\end{equation}
for all vector fields $X,Y$; but this can be rewritten
\begin{equation}
  g(A_\xi X, Y) = - g(X, A_\xi Y)~,
\end{equation}
which shows that $A_\xi$ is skew-symmetric relative to $g$.

In local coordinates, the Killing vector condition is
\begin{equation}
\nabla_M \xi_N + \nabla_N \xi_M = 0~,
\end{equation}
and the components of $A_\xi$, when thought of as a two-form, are
\begin{equation}
 A_{MN} = \nabla_M \xi_N = - \nabla_N \xi_M~.
\end{equation}
The skew-symmetric endomorphisms define a sub-bundle $\fso(TM) \subset
\End(TM)$.  Therefore a Killing vector $\xi$ gives rise to a section
$(\xi,A_\xi)$ of the bundle
\begin{equation}
  \label{eq:euclidean}
  \eE = TM \oplus \fso(TM)~.
\end{equation}
This can be understood as the local decomposition of a Killing vector
into a ``translation'' and a ``rotation''.  Of course, given the
Killing vector field $\xi$, $A_\xi$ is redundant as it can be
constructed from $\xi$. The importance of $A_\xi$ arises from the fact
that the Killing vector $\xi$ is completely determined by specifying
$(\xi(p),A_\xi(p))$ at a single point $p\in M$.

This is a consequence of the Killing identity which says that, for a
Killing vector $\xi$ and for any vector field $X$,
\begin{equation}
  \nabla_X A_\xi = R(X,\xi)~,
\end{equation}
where $R(X,Y)$ is the curvature operator defined by
\begin{equation}
  R(X,Y) = \nabla_X \nabla_Y - \nabla_Y \nabla_X - \nabla_{[X,Y]}~.
\end{equation}
This Killing identity\footnote{This identity can be proven as follows.
Take the covariant derivative of $A_\xi$ and use the
algebraic Bianchi identity to conclude that
\begin{equation}
   B(X,Y,Z) := g((\nabla_X A_\xi)Y,Z) - g(R(X,\xi)Y,Z)
\end{equation}
is symmetric in the first two entries.  It is also skew-symmetric in
the last two entries because $A_\xi$ (and hence $\nabla_X A_\xi$) is
skew-symmetric.  These two properties now imply that $B$ vanishes,
resulting in the Killing identity.}, which in local coordinates reads
\begin{equation}
\nabla_L \nabla_M \xi_N = R^P_{\;LMN}\xi_P~,
\end{equation}
($R^P_{\;LMN}$ are the components of the Riemann curvature tensor)
shows that indeed second and higher derivatives of $\xi$ at some point
$p\in M$ can be expressed recursively in terms of $\xi_M(p)$ and
$(\nabla_M \xi_N)(p)$.

In other words, the Killing identity together with the definition of
$A_\xi$ defines a differential system
\begin{equation}
  \nabla_X \xi - A_\xi X = 0 \qquad \text{and} \qquad
  \nabla_X A_\xi - R(X,\xi) = 0~,
\end{equation}
whose solutions can be interpreted as parallel sections of a suitable
connection on $\eE$.  Indeed, if $(\xi,A)$ is \emph{any} section of
$\eE$ we define its covariant derivative by
\begin{equation}
  \label{eq:Dconnection}
  D_X \xi = \nabla_X \xi - A(X) \qquad \text{and} \qquad
  D_X A = \nabla_X A - R(X,\xi)~.
\end{equation}
It follows that the parallel sections are precisely the sections
$(\xi,A)$ where $\xi$ is a Killing vector and $A = A_\xi$.

This means that a Killing vector $\xi$ is uniquely specified by the
value of $(\xi(p),A_\xi(p))$ at a point $p$, say, with the value at
any other point $q$ being determined by parallel transport along any
curve joining $p$ and $q$. (Recall that $M$ is assumed to be
connected.)  The value thus obtained is independent on the curve
because for a Killing vector, $(\xi, A_\xi)$ is invariant under
parallel transport around closed loops.  Incidentally, this explains
why the dimension of the isometry algebra of a $D$-dimensional
spacetime is at most $D(D+1)/2$, which is the rank of the bundle
$\eE$.

\subsubsection{Isometries are hereditary}

Let $(M,g)$ be a spacetime with $n$ linearly independent Killing
vectors.  We had already seen that for $\Omega>0$,
$(M_\Omega,g_\Omega)$ also has $n$ Killing vectors and the question is
what happens as $\Omega \to 0$.  We will use the local
characterisation of Killing vectors as parallel sections of the bundle
$\eE$ in \eqref{eq:euclidean} in order to apply the results of
Section~\ref{sec:geroch}.

To this end, let $\eE_\Omega = TM_\Omega \oplus \fso(TM_\Omega)$ and
let $D^\Omega$ be the connection defined by equation
\eqref{eq:Dconnection}.  Notice that the bundle and the connection
depend only on the metric.  Since the limit $\Omega \to 0$ of the
metric exists, so do the limits $\eE_0$ and $D^0$ of the bundle and
the connection.  Moreover, $D^0$ is the connection defined by
\eqref{eq:Dconnection} relative to the Penrose limit metric.  This is
the hypothesis which allows us to apply the argument in
Section~\ref{sec:geroch} to immediately conclude that in the Penrose
limit the number of linearly independent Killing vectors cannot
decrease and that these $n$ Killing vectors of $(M,\bar g = g_0)$
arise as limits of Killing vectors of $(M,g)$.

As mentioned above, the infinitesimal symmetries of a supergravity
background $(M,g,\Psi)$ are given by Killing vectors which in addition
leave invariant the other fields in the background.  These conditions
translate into linear equations $L_\xi\Psi=0$ on the Killing vector $\xi$.
Since the fields $\Psi$ have well-defined Penrose--Güven limits, these
equations are also well-defined in the limit, and using either the
argument of Section~\ref{sec:isometries1} or the (refined) argument
in Section~\ref{sec:geroch}, we conclude that the dimension of the
symmetry algebra of a supergravity background does not decrease in the
Penrose limit.

\subsection{Supersymmetries}
\label{sec:lordsusy}

We have shown that the Penrose limit of a solution of the supergravity
field equations is again a solution and moreover that it admits at
least as many symmetries as the original solution.  It is natural to
ask whether the limiting solution preserves at least as many
supersymmetries as the original solution.   The equations of a bosonic
supergravity background being supersymmetric translates into equations
on spinors coming from the supersymmetry variations of the fermions in
the theory.  The variation of the gravitini yields a Killing spinor
equation, whereas the variation of the other fermions in the theory
yield additional algebraic conditions.  As a result the above question
is essentially equivalent to asking whether the property of a spinor
being Killing is hereditary.

To answer this question, we can appeal to the generalisation of
Geroch's argument presented in Section~\ref{sec:geroch}.  The key
point is that the condition of being Killing can be interpreted as
the condition of being parallel with respect to a connection on the
spinor bundle $\spb$.  This again implies that any Killing spinor is
uniquely specified by its value at one point $p$ in spacetime.
Parallel transport will then define the Killing spinor everywhere.
There may be less supersymmetries than parallel spinors because of the
presence of the additional algebraic conditions, but the important
point is that the supersymmetries preserved by a bosonic background
are uniquely specified by their values at any one point in spacetime.

\subsubsection{Supersymmetries are hereditary}

We will first prove that the dimension of the space of Killing spinors
does not decrease in the Penrose limit.  This will follow once again
from the argument in Section~\ref{sec:geroch}.  Then we will show that
this conclusion does not change when we incorporate the algebraic
equations.

Let $(M,g)$ be a spacetime with $n$ linearly independent Killing
spinors.  It is clear that $(M_\Omega, g_\Omega,\Psi_\Omega)$ also
admits $n$ Killing spinors for all $\Omega>0$.  The question is
whether this persists in the limit $\Omega\to 0$. Let
$\varepsilon_{\Omega}$ be a Killing spinor of $(M_\Omega,
g_\Omega,\Psi_\Omega)$.  Then by linearity of the Killing spinor
equation also $\Omega^{\Delta}\varepsilon_{\Omega}$ is a Killing
spinor for any $\Delta$. As for Killing vectors, we can always find a
$\Delta$ such that the limit $\lim_{\Omega\to
  0}\Omega^{\Delta}\varepsilon_{\Omega}$ exists and is non-zero (by
choosing $\Delta$ to pick out the first non-zero coefficient in a
Taylor expansion around $\Omega=0$). Different Killing spinors may
require different $\Delta$'s, and this can lead to a contraction of
the superalgebra. To show linear independence of the Killing spinors
in the limit, we adapt Geroch's argument to the case at hand.

The spinor bundle $\spb$ depends on the metric,\footnote{It also depends
on a choice of spin structure.  However the limiting spacetime
  $M_0$, being homeomorphic to a neighbourhood of a segment of a null
  geodesic, is contractible and hence has a unique spin structure.
  More generally, the existence of a spin structure is hereditary and
  spin structures form a discrete set, so they cannot
  change continuously as we vary the parameter $\Omega$.}  and hence
defines a family $\spb_\Omega \to M_\Omega$ of spinor bundles.  The
supercovariant derivative depends on the metric and the other bosonic
fields in the background, and through this dependence it defines a
connection $D^\Omega$ on $\spb_\Omega$.  This family of bundles with
connection extends to the boundary $M_0$, since the metric and the
other bosonic fields in the supergravity theory have well-defined
limits as $\Omega\to 0$.  We can therefore appeal to the argument in
Section~\ref{sec:geroch} to conclude that there are at least as many
parallel sections of $\spb_0$ as there are of $\spb_\Omega$ for
$\Omega>0$.

Supersymmetries of a supergravity background are in one-to-one
correspondence with Killing spinors satisfying additional algebraic
linear equations, coming from the supersymmetry variations of the
other fermionic fields in the supergravity theory besides the
gravitini.  These algebraic equations depend on the bosonic fields in
the theory and since they have a well-defined Penrose--Güven limit,
so do the equations.  This allows us to apply the refinement of
Geroch's argument in Section~\ref{sec:geroch} to conclude that
the Penrose limit of a supergravity background $(M,g,\Psi)$ preserves
at least as many supersymmetries as $(M,g,\Psi)$.

It should be remarked that it is crucial in this assertion that the
limit $D^0$ of the supercovariant derivative is the supercovariant
derivative evaluated at the Penrose limit.  In other words, if we
write the dependence on the bosonic fields explicitly as
$D(g,\Psi)$, then we have that in the limit
\begin{equation}
  D^0 = D(\bar g,\bar \Psi)~.
\end{equation}

Finally let us mention that even if the original background is not
supersymmetric, its Penrose limits always preserve at least one-half
of the supersymmetry.  Indeed, the algebraic equations are satisfied
on the sub-bundle $\spb_-$ of spinors annihilated by $\Gamma_+$ because
of the form \eqref{eq:PLFSBrink} of the field-strengths in the Penrose
limit.  Furthermore, it is easy to show that on the sub-bundle
$\spb_-$ the connection $D^0$ defined above is flat.

\subsubsection{Eleven-dimensional supergravity}

Let us illustrate the above results with the example of
eleven-dimensional supergravity.  In some conventions, the
supercovariant derivative of eleven-dimensional supergravity theory is
given by
\begin{equation}
  \label{eq:superconn}
  D_M= \nabla_M - \Omega_M ~,
\end{equation}
where
\begin{equation}
  \label{eq:Omega}
  \Omega_M = \tfrac1{288}  F_{PQRS}
  \left( \Gamma^{PQRS}{}_M + 8 \Gamma^{PQR}\delta^S_M \right)~,
\end{equation}
and the spin connection $\nabla$ is
\begin{equation}
  \label{eq:spinconn}
  \nabla_M = \d_M + \tfrac14 \omega_M{}^{ab} \Gamma_{ab}~.
\end{equation}

Suppose now that we have an eleven-dimensional background which
preserves some supersymmetry, i.e., there are non-vanishing spinors
$\varepsilon$ such that
\begin{equation}
  \label{eq:Killing}
  D_M\varepsilon=0 ~.
\end{equation}
Clearly, this is a parallel transport equation of the spinor bundle
$\spb$ of eleven-dimensional spacetime $(M,g,F)$ of rank $32$.

Next observe that the above Killing spinor equation is well-defined at
a Penrose limit.  For this, let us adopt coordinates in the
neighbourhood of a null geodesic, scale the coordinates as described
in section \ref{sec:PGLimit} and perform the overall scaling $g\sto
\Omega^{-2} g$ and $A_3 \sto \Omega^{-3} A_3$ where $A_3$ is the
three-form gauge potential appropriately gauge-fixed.  Note that the
four-form field strength $F_4=dA_3$ is scaled as $F_4\sto
\Omega^{-3}F_4$.  Next we notice that the $\Gamma$ matrices in the
frame indices do not scale, whereas the eleven-dimensional frame
scales as $e_M^a\sto \Omega^{-1} e^a_M$.  Under these scalings, the
supercovariant derivative does not rescale: $ D_M \sto D_M$.
Therefore, the Penrose limit of the Killing spinor equation
\eqref{eq:Killing} is the standard eleven-dimensional supergravity
Killing spinor equation evaluated at the Penrose limit of the
associated spacetime.  Having established this, we can use the
argument of the previous section to show that any Penrose limit of a
supersymmetric solution of eleven-dimensional supergravity is
supersymmetric and it admits at least as many Killing spinors as the
original solution. In fact, it is easy to see that the Penrose limit
of any solution, even if the solution does not preserve any
supersymmetry, preserves at least sixteen supersymmetries.  This is
because all the pp-wave type solutions in eleven dimensions preserve
at least half the supersymmetry. The supersymmetry projection is
$\Gamma_+\varepsilon=0$.

\section{Penrose limits of $\AdS \times S$}
\label{sec:NHPL}

In this section we classify all Penrose limits of space-time
geometries of the form $\AdS \times S$.  As shown in
Section~\ref{sec:lordriem}, any Penrose limit of such a geometry is
locally isometric to a product of an indecomposable Cahen--Wallach
space with flat space.  More precisely, we will now prove that the
Penrose limit along any null geodesic in $\AdS_{p+2} \times S^n$ is
either flat or a Cahen--Wallach metric with two negative eigenvalues
in equal ratio to the radii of curvatures of the two factor spaces,
depending on whether or not the tangent component to the sphere of
geodetic vector vanishes.

In particular, we exhibit the maximally supersymmetric Hpp-wave
solutions to eleven-dimensional and IIB supergravity as Penrose limits
of the near horizon geometries of the M2/5 and D3 branes respectively.
We then discuss some generalisations of this construction and make
some comments about the fate of isometries under the Penrose limit.

\subsection{The $\AdS \times S$ metrics}
\label{sec:metrics}

We identify anti-de~Sitter space $\AdS_{p+2}$ with radius of curvature
$R_{\AdS}$ with the following quadric in the pseudo-euclidean space
$\EE^{2,p+1}$
\begin{equation}
  \label{eq:AdSquadric}
  (X^0)^2 + (X^{p+2})^2 - (X^1)^2 - \dots - (X^{p+1})^2 = R_{\AdS}^2
\end{equation}
with the induced metric.  Introduce the following parametrisation
\begin{equation}
  \label{eq:AdSparam}
  \begin{aligned}[m]
    X^0 &= R_{\AdS} \cos \tau\\
    X^{p+2} &= R_{\AdS} \sin \tau \sqrt{1 + r^2}\\
    X^i &= R_{\AdS} r \sin\tau \theta^i\quad\text{for $i=1,\dots,p+1$,}
  \end{aligned}
\end{equation}
where $\sum_i (\theta^i)^2 = 1$ parametrise a $p$-dimensional sphere.
In these coordinates, the anti-de~Sitter metric becomes
\begin{equation}
  \label{eq:ads}
  g_{\AdS} = R_{\AdS}^2 \left[-d\tau^2 + (\sin\tau)^2 \left(
  \frac{dr^2}{1+r^2} + r^2 d\Omega^2_{p}\right)\right]~,
\end{equation}
where $d\Omega^2_{p}$ is the $p$-sphere metric.

Similarly we identify the round $n$-sphere $S^n$ with radius of
curvature $R_S$ with the quadratic in ($n+1$)-dimensional euclidean
space $\EE^{n+1}$
\begin{equation}
  \label{eq:Squadric}
  (X^1)^2 + (X^2)^2 + \dots + (X^{n+1})^2 = R_S^2
\end{equation}
with the induced metric.  Let $\psi$ be the colatitude and write
\begin{equation}
  \label{eq:Sparam}
  \begin{aligned}[m]
    X^{n+1} &= R_S \cos \psi\\
    X^i &= R_S \sin\psi \omega^i\quad\text{for $i=1,\dots,n$,}
  \end{aligned}
\end{equation}
where $\sum_i (\omega^i)^2 = 1$ parametrise the equatorial
$(n-1)$-sphere.  In these coordinates, the round metric on the
$n$-sphere becomes
\begin{equation}
  \label{eq:sphere}
  g_S = R_S^2 \left[d\psi^2 + (\sin\psi)^2 d\Omega^2_{n-1}\right]~,
\end{equation}
where $d\Omega^2_{n-1}$ is the metric on the equatorial
($n-1$)-sphere.

The metric on $\AdS_{p+2} \times S^{D-p-2}$ is then
$g = g_{\AdS} + g_S$, which is given by
\begin{multline}
  \label{eq:adsxsmetric}
  R^{-2} g = \rho^2 \left[-d\tau^2 + (\sin\tau)^2 \left(
  \frac{dr^2}{1+r^2} + r^2 d\Omega^2_{p}\right)\right]\\
   + d\psi^2 + (\sin\psi)^2 d\Omega^2_{D-p-3}~,
\end{multline}
where we have introduced the ratio
$\rho := R_{\AdS_{p+2}}/R_{S^{D-p-2}}$ of the radii of curvature of
the two factors  and where $R$ is the radius of curvature of the sphere.

In particular, the near horizon geometries of the M2-, M5- and
D3-brane solutions are of the form $\AdS_p \times S^{D-p-2}$ where the
values of $p$ and $D$ corresponding to each of the above branes are
listed in Table~\ref{tab:pDradii} along with the ratio $\rho$.

\begin{table}[h!]
  \begin{center}
    \setlength{\extrarowheight}{5pt}
    \begin{tabular}{|c|c|c|c|}
      \hline
      Brane & $p$ & $D$ & $\rho$\\[3pt]
      \hline
      M2 & 2 & 11 & $\half$ \\
      D3 & 3 & 10 & 1 \\
      M5 & 5 & 11 & 2\\
      \hline
    \end{tabular}
    \vspace{8pt}
    \caption{Dimensions and radii of curvature}
    \label{tab:pDradii}
  \end{center}
\end{table}

\subsection{Penrose limits of $\AdS$}
\label{sec:PLads}

As a first step towards studying the Penrose limits of the $\AdS\times
S$ geometries in general, we determine the Penrose limits of $\AdS$
space-times. In fact, we will see that any Penrose limit of $\AdS$ is
flat Minkowski space. This could be deduced as a by-product of our
more general calculations below, establishing this result for
particular null geodesics, and then extended to all null geodesics
using the maximal symmetry of $\AdS$ and the covariance property of
Penrose limits.

However, there is also a simpler and more general argument requiring
no calculation and using only the hereditary properties of Penrose
limits. Indeed, we had seen in Section~\ref{sec:lordriem} that the
Penrose limit of any Einstein manifold is Ricci-flat. We had also seen
that the Penrose limit of a conformally flat space-time (vanishing
Weyl tensor) is necessarily conformally flat.

In particular, therefore, the Penrose limit of a maximally symmetric
(conformally flat, Einstein) manifold, either de~Sitter or
anti-de~Sitter space-time, has vanishing Ricci and Weyl tensors. This
implies that the Riemann curvature tensor is zero and hence that the
Penrose limit is isometric to Minkowski space-time.

\subsection{Classification of Penrose limits of $\AdS\times S$}
\label{sec:classPL}

Because of the covariance property of Penrose limits, in order to
classify the possible Penrose limits of these space-times we must
investigate the orbits of the isometry group $G$ of $M=\AdS_{p+2}
\times S^n$ acting on the space of pairs $(\gamma(0),[\Dot\gamma(0)])$
consisting of a point in $M$ and a future-pointing null direction at
that point.  Because $M$ is homogeneous,
\begin{equation}
M=  \frac{G}{H} = \frac{\SO(2,p+1) \times \SO(n+1)}{\SO(1,p+1) \times
  \SO(n)}~,
\end{equation}
the isometry group $G$ acts transitively on points. Once we have fixed
a point $\gamma(0)$, the subgroup fixing that point is isomorphic to
the stabiliser $H$ and it remains to investigate the action of $H$ on
the celestial sphere of null directions at a point.

It is easy to see that there are two orbits.  The ``big'' orbit is
$B^{p+1} \times S^{n-1}$, where the ball $B^{p+1}$ is the space of
future-pointing time-like directions at a point in $\AdS_{p+2}$ and
the sphere $S^{n-1}$ is the space of directions at a point in $S^n$.
There is also a smaller orbit diffeomorphic to $S^p$, the celestial
sphere at a point in $\AdS_{p+2}$, corresponding to null geodesics
$\gamma$ for which the component of $\Dot\gamma$ tangent to $S^n$
vanishes.

By the covariance property there are then at most two non-isometric
Penrose limits corresponding to the two orbits. We will show that the
small orbit gives rise to a flat Penrose limit - this is almost
obvious from the fact we established above that the Penrose limit of
pure $\AdS$ is flat. We will then show that the Penrose limit
corresponding to the large orbit (which is hence generic) gives rises
to a Cahen--Wallach spacetime where the matrix $A_{ij}$ has two
negative eigenvalues with multiplicities $p+1$ and $n-1$,
respectively, commensurate with the radii of curvature of $\AdS_{p+2}$
and $S^n$. In particular, this will establish that the Penrose limits
of the near horizon geometries of the D3-, M2-, and M5-branes are the
maximally supersymmetric Hpp-waves of 10- and 11-dimensional
supergravity.

\subsection{The non-generic orbit: null geodesics in $\AdS$}
\label{sec:PLadss1}

We consider null geodesics of $\AdS\times S$ tangent to $\AdS$.
Because of this and because the metric is the product metric, we can
investigate the fate of the two factors separately. We had already
seen above that the Penrose limit of the $\AdS$ factor is flat. It
thus remains to consider what happens to the metric on the sphere $S$
in the Penrose limit. We know from Section~\ref{sec:PLblowup} that one
consequence of the Penrose limit is to blow up the geodesic distance
between any two points on $S$. As in the present case this blowing up
is uniform on all of $S$, clearly this implies that in the Penrose
limit the metric on $S$ is the infinite radius flat metric.

One can also see this directly in local coordinates. Let the
line-element on the sphere $S$ be
\begin{equation}
  g_{S}=g_{ab}(Y^{c})dY^{a}dY^{b}~.
\end{equation}
As the $Y^{a}$ are among the transverse coordinates $Y^{i}$, in the
Penrose limit they scale as (see \eqref{eq:diffeo}) $Y^{a} = \Omega
y^{a}$ so that
\begin{equation}
  g_{S}(\Omega) = \Omega^{2} g_{ab}(\Omega y^{c}) dy^{a} dy^{b}~.
\end{equation}
It thus follows from \eqref{eq:rescaling} that the metric in the
Penrose limit is the constant (hence flat) metric
\begin{equation}
  \bar g_{S} = \lim_{\Omega\to 0} \Omega^{-2} g_S(\Omega) =
  g_{ab}(0)dy^{a}dy^{b}~.
\end{equation}

Putting this together we deduce that the Penrose limit of $\AdS \times
S$ for any null geodesic in the non-generic orbit, i.e., tangent to
$\AdS$, is Minkowski space-time.

\subsection{Generic null geodesics}

We now consider null geodesics in $\AdS\times S$ with a non-zero
component along $S$. There are two ways to approach the calculation of
the Penrose limit. One is to solve the geodesic equation for a
suitable initial condition and then to find a coordinate
transformation which puts the metric into the form \eqref{eq:metric}
adapted to the null geodesic congruence. The other is to forego the
determination of a null geodesic and directly find a coordinate
transformation which puts the metric into the canonical form
\eqref{eq:metric} which then exhibits $\d/\d U$ as the null geodesic.
While ultimately both methods are equivalent, the former is
occasionally more transparent while the latter may be quicker. We will
illustrate both methods in the following, starting with the second
method.

We follow closely the discussion in \cite{ShortLimits} where the emphasis
was on the near horizon geometries
$\AdS_5 \times S^5$ and $\AdS_{4|7}\times S^{7|4}$.
Recall the $\AdS\times S$ metric \eqref{eq:adsxsmetric},
\begin{multline}
  R^{-2} g = \rho^2 \left[-d\tau^2 + (\sin\tau)^2 \left(
  \frac{dr^2}{1+r^2} + r^2 d\Omega^2_{p}\right)\right]\\
   + d\psi^2 + (\sin\psi)^2 d\Omega^2_{D-p-3}~.
\end{multline}

Let us now change coordinates in the $(\psi,\tau)$ plane to
\begin{equation}
  \label{eq:lightcone}
  u = \psi +\rho \tau \qquad  v = \psi -\rho \tau~,
\end{equation}
in terms of which, the metric $g$ becomes
\begin{multline}
  R^{-2} g = du dv + \rho^2 \sin((u-v)/2\rho)^2 \left(
    \frac{dr^2}{1+r^2} + r^2 d\Omega^2_{p}\right)\\
  + \sin((u+v)/2)^2 d\Omega^2_{D-p-3}~.
\end{multline}
We now take the Penrose limit along the null geodesic parametrised by
$u$.  In practice this consists in dropping the dependence on other
coordinates but $u$.  Doing so we find
\begin{equation}
  \label{eq:PLRosen}
  R^{-2} \bar g = du dv + \rho^2 \sin(u/2\rho)^2
  ds^2(\EE^{p+1}) + (\sin (u/2))^2 ds^2(\EE^{D-p-3})~,
\end{equation}
which is, as we have seen in Section~\ref{sec:PLCW},
the metric of a Cahen--Wallach symmetric space in Rosen
coordinates (compare with \cite{CKG} for the $d=11$ solution).

To determine the resulting Cahen--Wallach space more explicitly,
let us introduce coordinates $y^a$ for $a=1,\dots,D-2$ in
such a way that the metric \eqref{eq:PLRosen} becomes
\begin{equation}
  R^{-2} \bar g = du dv + \sum_{a=1}^{D-2} \frac{(\sin\lambda_a
  u)^2}{(2\lambda_a)^2} dy^a dy^a~,
\end{equation}
where
\begin{equation}
  \label{eq:lambdas}
  \lambda_a =
  \begin{cases}
    1/2\rho & a=1,\dots,p+1\\
    1/2 & a=p+2,\dots,D-2~.
  \end{cases}
\end{equation}
We change coordinates to $(x^+, x^-, x^a)$ where
\begin{equation}
  \label{eq:BrinkmanCoords}
  x^- = u/2~, \quad x^+ = v - \tfrac14 \sum_a y^a y^a
  \frac{\sin(2\lambda_a u)}{2\lambda_a}~, \quad x^a= y^a
  \frac{\sin(\lambda_a u)}{2\lambda_a}~,
\end{equation}
so that the metric now becomes
\begin{equation}
  \label{eq:PLCW}
  R^{-2} \bar g = 2 dx^+ dx^- - 4\left( \sum_a \lambda_a^2 x^a x^a
  \right) (dx^-)^2   + \sum_a dx^a dx^a~,
\end{equation}
which we recognise as a Cahen--Wallach metric \eqref{eq:CWtype} whose
matrix $A_{ij}$ is constant and diagonal with negative eigenvalues
$\{-\lambda_a^2\}$.  For $\lambda_a$ given as in \eqref{eq:lambdas} we
obtain \cite{ShortLimits}, if $\rho=\half$ or $\rho=2$, precisely the
metrics of the maximally supersymmetric Hpp-waves of
eleven-dimensional supergravity \eqref{eq:11dHpp} discovered in
\cite{KG} (see also \cite{FOPflux}), and if $\rho=1$ the maximally
supersymmetric Hpp-wave of IIB supergravity \eqref{eq:IIBHpp}
discovered in \cite{NewIIB}.  In \cite{ShortLimits} it was also shown
that the limits of the corresponding ($D-p-2$)-form field strengths
agree with those of the Hpp-waves \cite{ShortLimits}.

We conclude that the maximally supersymmetric Hpp-waves of
\cite{KG,FOPflux} and \cite{NewIIB} appear as Penrose limits along
generic null geodesics of the near horizon geometries of the M2/5 and
D3 branes, respectively \cite{ShortLimits}. Likewise, the maximally
supersymmetric Hpp-waves in four \cite{KGd4} and five and six
dimensions \cite{Meessen} appear as Penrose limits of the $AdS_2\times
S^2$, $\AdS_2\times S^3$ or $\AdS_3\times S^2$, and $\AdS_3\times S^3$
solutions of the corresponding supergravity theories. The latter is
the near-horizon limit of the six-dimensional self-dual string
\cite{DuffLuSDS} whose Penrose limits we will discuss in
Section~\ref{sec:sdstring}.

In the light of the results of section~\ref{sec:lordsusy}, this
derivation also provides an alternative proof that these solutions are
indeed maximally supersymmetric.

Moreover we now understand the (originally somewhat puzzling) fact
that the IIB Hpp-wave is characterised by a matrix $A_{ij}$ with a
single eigenvalue with multiplicity eight, whereas the Hpp-wave of
eleven-dimensional supergravity has two distinct eigenvalues with
multiplicities $3$ and $6$ respectively. It is is related to the fact
that the two curvature radii of the IIB $\AdS \times S$ solution are
equal whereas those of the solutions of eleven-dimensional
supergravity are not.

We also see that we can obtain any indecomposable Cahen--Wallach
metric with $A_{ij}$ having at most two (negative) eigenvalues as the
Penrose limit of a product $\AdS_m \times S^n$ by appropriate choices
of $m$, $n$ and the ratio of radii of curvature of the two factors.
More generally, other Cahen--Wallach metrics with multiple eigenvalues
of any sign can be obtained as the Penrose limit of products involving
one (anti) de Sitter space and multiple spheres and hyperbolic spaces
of appropriate dimensions and radii of curvature.

\subsection{Isometries revisited}
\label{sec:isometriestoo}

In this section we illustrate the discussion in
Sections~\ref{sec:isometries1} and \ref{sec:isometries2} about the
fate of isometries under the Penrose limit with some examples of the
form $\AdS \times S$.

We start with a ``toy model'' corresponding to the near horizon
geometry of the Reissner--Nordström black hole in four-dimensional
$N{=}2$ supergravity, namely $\AdS_2 \times S^2$ with equal radii of
curvature.  The $\SO(2,1) \times \SO(3)$ isometry group of this space
is most easily exhibited by embedding $\AdS_2 \times S^2$ as the
intersection of two quadrics in $\EE^{2,4}$.  This is the case $p=0$,
$D=4$ (and hence $n=2$) in the notation of Section~\ref{sec:metrics}.
An explicit parametrisation is given by
\begin{equation}
  \label{eq:AdS2xS2emb}
  \begin{split}
    X^0 = R\cos\tau~,\quad X^1=R \sinh\beta \sin\tau
    \quad\text{and}\quad X^2 = R \cosh\beta \sin\tau\\
    X^3 = R\cos\psi~,\quad X^4=R \sin\psi\cos\theta
    \quad\text{and}\quad X^5 = R \sin\psi \sin\theta~,
  \end{split}
\end{equation}
where $R$ is the common radius of curvature of the two spaces.  In
terms of the ambient coordinates, a basis for Killing vectors of $\AdS_2
\times S^2$ is given by the six vector fields
\begin{equation}
  \begin{split}
    X^1 \d_i + X^i \d_1 \qquad\text{for $i=0,2$, and}\\
    X^i \d_j - X^j \d_i \qquad\text{for $i,j=3,4,5$ and $i,j=0,2$,}
  \end{split}
\end{equation}
where $\d_i = \d/\d X^i$.  In terms of the embedding coordinates,
these Killing vectors are explicitly given by
\begin{equation}
  \begin{aligned}[t]
    E_1 &=  \frac{\d}{\d\theta}\\
    E_1^* &=  \sin\theta \frac{\d}{\d\psi} + \cos\theta \cot\psi
    \frac{\d}{\d\theta}\\
    \xi_1 &= \cos\theta \frac{\d}{\d\psi} - \sin\theta \cot \psi
    \frac{\d}{\d\theta}
  \end{aligned}
  \qquad
  \begin{aligned}[t]
    E_2&=  \frac{\d}{\d\beta}\\
    E_2^*&=  -\sinh\beta \frac{\d}{\d\tau} + \cosh\beta \cot\tau
    \frac{\d}{\d\beta}\\
    \xi_2 &= -\cosh\beta \frac{\d}{\d\tau} + \sinh\beta \cot \tau
    \frac{\d}{\d\beta}
  \end{aligned}
\end{equation}
The Penrose limit starts by rescaling the coordinates as follows
\begin{equation}
  \label{eq:AdS2xS2resc}
  \psi = \half (u + \Omega^2 v)~,\quad \tau = \half( u - \Omega^2 v)~,
  \quad \theta = \Omega y^1 \quad\text{and}\quad \beta = \Omega y^2~.
\end{equation}
In terms of the new variables, the Killing vectors acquire $\Omega$
dependence.  To leading order in $\Omega$ one finds, respectively,
\begin{equation}
  \begin{aligned}[t]
    E_1(\Omega)&=\Omega^{-1} \frac{\d}{\d y^1}\\
    E_1^*(\Omega)&= \Omega^{-1}\left(y^1 \frac{\d}{\d v} + \cot\half u
      \frac{\d}{\d y^1}\right)\\
    \xi_1(\Omega)&= \Omega^{-2}\frac{\d}{\d v}
  \end{aligned}
  \quad
  \begin{aligned}[t]
    E_2(\Omega)&=  \Omega^{-1} \frac{\d}{\d y^2}\\
    E_2^*(\Omega)&= \Omega^{-1}\left(y^2 \frac{\d}{\d v} + \cot\half u
    \frac{\d}{\d y^2}\right)\\
  \xi_2(\Omega)&=  \Omega^{-2}\frac{\d}{\d v}
  \end{aligned}
\end{equation}
which are to be compared with the expression \eqref{eq:CWisomRosen}
for the generic isometries of a Penrose limit.

We observe the phenomena described in Section~\ref{sec:isometries1}.
First of all, we see that the rescaled Killing vectors
\begin{equation}
  \Omega E_i(\Omega) \qquad \Omega E_i^*(\Omega) \qquad
  \Omega^2\xi_i(\Omega)
\end{equation}
have well-defined limits as $\Omega\to 0$. Secondly, we notice that
two linearly independent Killing vectors, namely $\xi_1(\Omega)$ and
$\xi_2(\Omega)$ are equal up to subleading terms in $\Omega$.  Since
we know that we cannot lose any Killing vectors in the limit, we are
thus led to consider the linear combinations
\begin{equation}
  \xi_\pm(\Omega) = \xi_1(\Omega) \pm \xi_2(\Omega)~.
\end{equation}
To leading order in $\Omega$ one has
\begin{align}
  \xi_+(\Omega) &= 2\Omega^{-2} \frac{\d}{\d v}\\
  \xi_-(\Omega) &= 2 \frac{\d}{\d u} - \half |y|^2 \frac{\d}{\d v} -
  \sum_i y^i \cot\half u \frac{\d}{\d y^i}~,
\end{align}
so that $\Omega^2 \xi_+(\Omega)$ and $\xi_-(\Omega)$ are well defined
(and linearly independent) in the limit. Comparing with
\eqref{eq:CWisomRosen}, we see that
\begin{equation}
  e_i :=\lim_{\Omega\to 0} \Omega E_i(\Omega)\qquad
  e_i^* :=\lim_{\Omega\to 0} \Omega E_i^*(\Omega)\qquad
  e_+ :=\tfrac12\lim_{\Omega\to 0}\Omega^{2}\xi_+(\Omega)
\end{equation}
are precisely the Killing vectors of a generic Penrose limit pp-wave
spacetime satisfying the Heisenberg algebra \eqref{eq:heisenberg},
only that in this case they arise from isometries already present in
the original spacetime.

Moreover, in this case, since the Penrose limit is actually an
Hpp-wave, there is another Killing vector, namely $\d/\d x^-$ in
Brinkman coordinates (as $A_{ij}$ is constant, independent of $x^-$)
or
\begin{equation}
  \frac{\d}{\d x^-}=  2 \frac{\d}{\d u} - \half |y|^2 \frac{\d}{\d v}
  - \sum_i y^i \cot\half u \frac{\d}{\d y^i}
\end{equation}
in Rosen coordinates. We see that this agrees precisely with
\begin{equation}
e_-:=\lim_{\Omega\to 0}\xi_-(\Omega)~,
\end{equation}
so that this Killing vector is again inherited from an isometry of the
original $\AdS\times S$ spacetime.

We also see that the Killing vectors have to be rescaled by different
powers of $\Omega$ to have a well-defined limit, exactly as in the
rescaling \eqref{eq:diffeo} of the coordinates, by $\Omega$ for the
transverse directions and by $\Omega^0$ and $\Omega^2$ for the
directions corresponding to $\d/\d u$ and $\d/\d v$. Therefore the
isometry algebra will get contracted in the limit.  Here $\fso(1,2)$
and $\fso(3)$ both get contracted to Heisenberg algebras and
furthermore their central elements are identified in the limit to
become the common central element of the combined Heisenberg algebras.
$\xi_-(\Omega)$, on the other hand, becomes an outer automorphism of
the Heisenberg algebra.

Moreover in this case we have an additional isometry because of the
fact that the two spaces $\AdS_2$ and $S^2$ have equal radii of
curvature.  This accidental isometry manifests itself as rotations in
the $y^1,y^2$ plane: $y^1 \d/\d y^2 - y^2 \d/\d y^1$.  These extra
isometries commute with each other and act on the generic isometries
as outer automorphisms of the Heisenberg algebra.

More generally, we can consider the Penrose limit of $\AdS_{p+2}\times
S^{D{-}p{-}2}$ with radii of curvature $R_{\AdS}=\rho R$ and $R_S=R$,
respectively.  The Penrose limit along the null geodesic considered
above is a Cahen--Wallach metric where the (constant) matrix $A_{ij}$
has two eigenvalues with ratio $\rho$ and multiplicities $p+1$ and
$D{-}p{-}3$.  Under the Penrose limit, the isometry algebra
$\fso(2,p+1) \oplus \fso(D{-}p{-}1)$ of $\AdS_{p+2}\times
S^{D{-}p{-}2}$ undergoes the following contraction.  The $\fso(2,p+1)$
factor contracts to $\fh(p+1)\rtimes\fso(p+1)$, where $\fh(p+1)$ is a
Heisenberg algebra with $2p+3$ generators, whose $p+1$ creation and
$p+1$ annihilation operators transform as vectors under $\fso(p+1)$.
Similarly the $\fso(D{-}p{-}1)$ factor contracts to
$\fh(D{-}p{-}3)\rtimes\fso(D{-}p{-}3)$.  The central element in both
Heisenberg algebras coincide.  This means that there are two Killing
vectors $\xi_1(\Omega)$ and $\xi_2(\Omega)$ agreeing to leading order
in $\Omega$ and hence agreeing in the limit.  This prompts us to
consider the linear combinations $\xi_\pm(\Omega) = \xi_1(\Omega) \pm
\xi_2(\Omega)$.  These vector fields must be rescaled differently
for their limits to exist: $\xi_+(\Omega)$ becomes in the limit the
common central element of the combined Heisenberg algebra
$\fh(D{-}2)$, whereas $\xi_-(\Omega)$ becomes an outer automorphism
commuting with $\fso(p+1)\oplus\fso(D{-}p{-}3)$.  In Brinkman
coordinates, $\bar\xi_\pm$ are realised as $\d/\d x^\pm$.  We see,
therefore, that the isometry algebra $\fso(2,p+1) \oplus
\fso(D{-}p{-}1)$ of $\AdS_{p+2}\times S^{D{-}p{-}2}$ contracts to a
semidirect product
\begin{equation}
  \fh(D{-}2) \rtimes \left( \fso(p+1) \oplus \fso(D{-}p{-}3) \oplus
  \RR \right)~.
\end{equation}
When the radii of curvature are equal there is an additional
symmetry enhancement, and the subalgebra
$\fso(p+1)\oplus\fso(D{-}p{-}3)$ is enlarged to the full
$\fso(D{-}2)$.  This however has no counterpart in the original
metric.

\subsection{Generic null geodesics: another example}

Our general arguments on covariance of the Penrose limit, combined
with the above explicit calculations, now tell us that we know the
Penrose limit for any null geodesic in $\AdS \times S$. Nevertheless,
for illustrative purposes we will now look at another example which
allows us to see more explicitly what happens to the Penrose limit of
a generic geodesic as the angular momentum of the geodesic along the
sphere vanishes (and hence the geodesic approaches a non-generic
geodesic).

To be specific, we consider $\AdS_{5}\times S^{5}$ with $\AdS$ in
Poincaré coordinates times the sphere in standard spherical
coordinates,
\begin{equation}
  ds^2 = r^{-2}dr^2+ r^2 (-dt^2 + ds^2(\EE^3)) + d\psi^2 +
  (\sin\psi)^2 ds^2(S^{4}) ~,
\end{equation}
and look at null geodesics in the $(r,t,\psi)$-direction. Thus the
metric we will actually be working with is
\begin{equation}
  ds_{(3)}^2 = r^{-2}dr^2 -r^2 dt^2  + d\psi^2 ~.
\end{equation}
The null condition gives
\begin{equation}
  r^{-2}\dot{r}^2 +\dot{\psi}^2 = r^2 \dot{t}^2~.
\end{equation}
Energy and angular momentum conservation ($t$- and $\psi$-independence
of the metric) lead to
\begin{equation}
  r^2\dot{t} = E~, \qquad \dot{\psi} = \ell~.
\end{equation}
Hence one obtains
\begin{equation}
  \dot{r}^2 + \ell^2 r^2 = E^2~,
\end{equation}
which is solved by
\begin{equation}
  r(\tau) = \ell^{-1}E \sin \ell\tau~,
\end{equation}
and therefore
\begin{equation}
  t(\tau) = -E^{-1}\ell \cot \ell\tau \qquad\text{and}\qquad
  \psi(\tau) = \ell\tau~.
\end{equation}
Here without loss of generality we have set all integration constants
to zero (they will, in any case, reappear below as the transverse
coordinates parametrising the congruence of null geodesics).
Without loss of generality we can also choose $E=1$
by a rescaling of $\tau$.

Having obtained this congruence of null geodesics, the next step is to
change coordinates to an adapted coordinate system
\begin{equation*}
(r,t,\psi) \to (u,v,\phi)
\end{equation*}
where $u$ is the parameter $\tau$ along the null geodesics, i.e.,
$\d_{u}$ is the null geodesic vector field with $g_{uu} \equiv
g(\d_{u},\d_{u}) = 0$, and otherwise characterised by $g_{uv} =1$ and
$g_{u\phi} = 0$.  A possible choice is
\begin{equation}
  \begin{split}
    \d_{u} &= \dot{r}\d_{r} + \dot{t}\d_{t} + \dot{\psi}\d_{\psi}\\
    &= (1-\ell^{2}r^2)^{1/2}\d_{r} + r^{-2} \d_{t} + \ell\d_{\psi}\\
    \d_{v} &= - \d_{t}\\
    \d_{\phi}&= \d_{\psi} + \ell \d_{t}~.
  \end{split}
\end{equation}
This integrates to
\begin{equation}
  \begin{split}
    r(u,v,\phi) =& \ell^{-1}\sin \ell u\\
    t(u,v,\phi) =& -\ell \cot \ell u -v + \ell\phi\\
    \psi(u,v,\phi) =& \phi + \ell u~,
  \end{split}
\end{equation}
so that $(v,\phi)$ have the interpretation of constants of integration
in the geodesic equation parametrising the congruence of null
geodesics.

The next step is to express the metric in the new variables. By
construction, $g_{uu}=g_{u\phi} = 0$, and one finds
\begin{equation}
 \begin{split}
ds^2&=2 du dv + 2 \ell^{-1}(\sin \ell u)^2 dv\,d\phi - \ell^{-2}(\sin \ell u)^2 (dv)^{2}
+ (\cos \ell u)^2\; (d\phi)^{2}\\
 &+  \ell^{-2}(\sin \ell u)^2 ds^2(\EE^3) + (\sin(\ell u+\phi))^2 ds^{2}(S^4)~.
 \end{split}
\label{eq:ex2metric}
\end{equation}

We will now show that the Penrose limit of the above metric is flat
Minkowski space if and only if $\ell=0$ and the maximally
supersymmetric Hpp-wave of IIB supergravity otherwise.

In order to establish this, it is useful to recall the result
\eqref{eq:bi} on the form of $A_{ij}(x^-)$ for metrics with diagonal
$C_{ij}(u)=a_i(u)^2 \delta_{ij}$ in the Penrose limit in Rosen
coordinates.  We had seen that $A_{ij}=0$ (i.e., the Penrose limit is
flat) if and only if $a_{i}(u) = b_{i} + c_{i}u$.

Moreover, since for constant $A_{ij}$ we can always absorb a positive
constant $\mu^2$ into $A_{ij}$ by a scaling of $(x^+,x^-)$, we learn
that a metric in Rosen coordinates with diagonal $C_{ij}$ is
equivalent to the maximally supersymmetric IIB Hpp-wave
\eqref{eq:IIBHpp} if and only if
\begin{equation}
  a_{i}(u)'' = -\mu^2 a_{i}(u)
\end{equation}
for an $i$-independent non-zero constant $\mu$, i.e., if and only if
\begin{equation}
  a_{i}(u) = b_{i} \sin \mu u + c_{i} \cos \mu u~.
\end{equation}

Having established this, it is now easy to determine what is the
Penrose limit of the metric \eqref{eq:ex2metric}.  Clearly for
$\ell\neq 0$ the metric is of the required form (with $\mu=\ell$) to
give the Hpp wave solution: the second and third term disappear and
the remaining terms are of the required trigonometric form.

In the limit $\ell\to 0$, on the other hand, even before taking the Penrose
limit, the second (rotation) term goes to zero and the
 fourth and sixth term combine
to the line element on $\EE^5$. Thus the metric becomes
\begin{equation}
ds^2_{\ell=0} = 2 du\, dv + u^2 (-(dv)^2 + ds^2(\EE^3)) + ds^{2}(\EE^5)~,
\end{equation}
which is the ten-dimensional Minkowski spacetime in the Penrose limit.

\section{Penrose limits of branes}
\label{sec:branes}

In this and the next sections we will investigate the Penrose limits
of supergravity brane solutions.  We will limit our discussion in
this section to elementary $p$-brane solutions for which the isometry
group of the $D$-dimensional spacetime metric is $\ISO(1,p) \times
\SO(D{-}p{-}1)$.  Intersecting brane configurations will be discussed
in the next section.

\subsection{Classification of null geodesics}

As discussed above for the near horizon geometries, one way to
classify the different null geodesics, and hence the different Penrose
limits, is to use the covariance property, by which two null geodesics
which are related by an isometry induce isometric Penrose limits. We
therefore need to study the orbits of the isometry group on the
space of null geodesics, which is the space of pairs consisting of a
point in the space time (the ``initial'' point of the geodesic) and a
null direction at that point.

The typical metric for a supergravity brane solution in $D$
dimensions is
\begin{equation}
\label{eq:brane}
\begin{aligned}[m]
  ds^2&=A^2(r) ds^2(\EE^{(1,p)})+B^2(r) ds^2(\EE^{D{-}p{-}1})\\
  F_{p+2}&=\dvol(\EE^{1,p})\wedge dC(r)\\
  \phi&=\phi(r)
  \end{aligned}
\end{equation}
where $r$ is the radial transverse coordinate, $\EE^{(1,p)}$ is
the worldvolume of the brane and $\EE^{D{-}p{-}1}$ is the
transverse space. The components $A,B$ and $C$ as well as the
scalar $\phi$ for a single brane or for many branes located at the
same point depend only on the radial coordinate $r$.

Unlike the near horizon geometries, which are homogeneous, the
isometry group $G=\ISO(1,p) \times \SO(D{-}p{-}1)$ acts on the brane
solutions with cohomogeneity one.  The orbits are labelled by the
radial distance $r$ transverse to the brane: for $r>0$ they are
diffeomorphic to $\RR^{p+1} \times S^{D{-}p{-}2}$ and have codimension
one.  (As $r\to 0$ one recovers the near horizon geometry which, at
least for the M2, D3 and M5 branes, was already discussed above.)

The isotropy subgroup of a point $P$ a distance $r>0$ away from the
brane is isomorphic to $H = \SO(1,p) \times \SO(D{-}p{-}2)$.  To study
how this group acts on the null directions, we break the tangent space
$T_P M$ at $P$ to the spacetime manifold $(M,g)$ describing the brane
into three orthogonal subspaces
\begin{equation}
  T_P M = T_P B \oplus T_P R \oplus T_P S~,
\end{equation}
where $T_P B$ are those vectors tangent to the brane, $T_P R$ is the
radial component and $T_P S$ are those vectors tangent to the
transverse sphere.  The metrics on each of the factors depend only on
the radial distance $r$.  A null geodesic in $M$ is specified by the
initial point $P$ and by a null direction at $P$; that is, (the
projectivisation of) a null vector in $T_P M$.  Let $V$ be one such
vector and let $V = V_B + V_R + V_S$ be its decomposition relative to
the above splitting.  Notice that $V_R$ is determined up to a sign by
$V_B$ and $V_S$, since $V$ is null.  We must therefore distinguish two
cases:
\begin{enumerate}
\item \emph{$V_B$ is null; hence $V_R = V_S = 0$.}  Since $\SO(1,p)$
  acts transitively on the celestial sphere of $T_P B$, all geodesics
  starting at $P$ in the direction of $V$ are equivalent.
\item \emph{$V_B$ is timelike.}  We must distinguish between two
  subcases:
  \begin{enumerate}
  \item \emph{$V_S = 0$.}  $\SO(1,p)$ acts transitively on timelike
    directions in $T_P B$.  We simply normalise $V_B$ appropriately
    and $V_R$ is determined up to a sign; but this sign can be changed
    by changing the sign of the affine parameter along the geodesic
    and using a Lorentz transformation in $\SO(1,p)$ to reverse the
    sign of $V_B$.
  \item \emph{$V_S \neq 0$.}  $H$ acts transitively on timelike
    directions in $T_P B$ and on directions in $T_P S$; but now the
    relative scale matters.  Therefore there is a free parameter in
    this case: the ratio of the norms of $V_B$ and $V_S$.  Any two
    choices with the same ratio are equivalent under $H$.  Again, the
    radial component $V_R$ is determined up to a sign by the condition
    that $V$ is null.  As above, this sign is immaterial.
  \end{enumerate}
\end{enumerate}

We will call geodesics in 1 \emph{longitudinal} and those in 2(a)
\emph{radial}, whereas those in 2(b) are \emph{generic} and we will
now in turn discuss the associated Penrose limits.

\subsection{Longitudinal null geodesics}

Without loss of generality (covariance of the Penrose limit) we can choose
the longitudinal null geodesic to lie
in the $(t,x)$-plane where $x$ is any of the longitudinal
worldvolume coordinates. A coordinate system adapted to
such a null geodesic sitting at a fixed regular (non-zero)
 value $r_0$ of the transverse radial coordinate $r$ is
\begin{equation}
  \begin{aligned}[m]
    v&=x-t \\
    u&=\half (x+t)A(r)^2\\
    \rho&=r -r_0
 \end{aligned}
\end{equation}
with all the other coordinates unchanged. In terms of these coordinates, the
metric in \eqref{eq:brane} reads
\begin{equation}
 \begin{split}
  ds^2&=2 dudv - 4 A(\rho +r_0)'A(\rho+r_0)^{-3}udv d\rho\\
      &+ A^2(\rho+r_0) ds^2(\EE^{p-1})+B^2(\rho+r_0) ds^2(\EE^{D{-}p{-}1})
 \end{split}
\end{equation}
which is indeed of the required form \eqref{eq:metric}. In the Penrose limit
one finds
\begin{equation}
  ds^2=2 dudv + A^2(r_0) ds^2(\EE^{p-1})+B^2(r_0) ds^2(\EE^{D{-}p{-}1})~.
\end{equation}
This is isometric to the flat Minkowski metric on $\EE^{(1,D-1)}$ for
any $r_0 >0$. In addition the other fields associated with brane
solutions either vanish or become constant in the limit. In particular
the various form field strengths vanish and the scalars become
constant.

\subsection{Radial null geodesics}

We will now provide a fairly complete analysis of the Penrose limit of
brane solutions along radial null geodesics, dealing with D-branes,
M-branes and NS-branes. In this case explicit formulae
can be given for the various brane solutions in the limit.

\subsubsection{Some general remarks on radial null geodesics}

We can choose
the timelike component $V_B$ to lie in the time-direction. Thus
we shall investigate the Penrose limit involving the worldvolume time
coordinate $t$ and the radial transverse coordinate $r$.  We
shall focus mainly on the limit involving the metric. The Penrose
limit for the rest of the fields will be described at the end. To
achieve this, we first write the metric in \eqref{eq:brane} as
\begin{equation}\label{met}
ds^2=A^2(r) ds^2(\EE^{(1,p)})+B^2(r) \left( dr^2+r^2
ds^2(S^{D{-}p{-}2}) \right)~.
\end{equation}
To adapt coordinates appropriate for taking the Penrose limit we
have to find coordinates $(u,v)$ to rewrite the two-dimensional
metric
\begin{equation}
  ds_{(2)}^2=-A^2(r) dt^2+B^2(r) dr^2
\end{equation}
as
\begin{equation}
  ds^2=2dudv+ D^2(u,v) dv^2
\end{equation}
This implies that the vector $\d_{u}\equiv \d/\d u$ is null and geodetic.
For this we consider the coordinate transformations
\begin{equation}
  \begin{aligned}[m]
    v&=t+a(r) \\
    u&=-t+b(r)
 \end{aligned}
\end{equation}
Using these coordinate transformations and comparing the two
expressions for the metric in the two coordinate systems, we find
\begin{equation}
  \begin{aligned}[m]
    D^2&=2-A^2\\
    b'&=(A^2-1) a'\\
    (a')^2&=B^2/A^2
  \end{aligned}
\end{equation}
where ${}'$ denotes differentiation with respect to the coordinate
$r$. This can be rewritten as
\begin{equation}
  \begin{aligned}[m]
    D^2&=2-A^2\\
    b'&=\pm(A^2-1) B/A\\
    a'&=\pm B/A
  \end{aligned}
\end{equation}
In particular observe that $u+v=a+b$ and $(a+b)'=\pm AB$. This is
a key equation because it gives the transformation between the
$u+v$ coordinate and $r$ and so $r=r(u+v)$.

Next rewriting the brane metric in the above coordinate system,
we get
\begin{multline}
  ds^2= 2 du dv +D^2(u,v) dv^2+A^2\left(r(u+v)\right)
  ds^2(\EE^{p})\\
  + B^2\left(r(u+v)\right) r^2(u+v) ds^2(S^{D{-}p{-}2})
\end{multline}
Taking the Penrose limit, we find
\begin{equation}
  ds^2= 2 du dv + A^2 \left( r(u)\right) ds^2(\EE^p) + B^2
  \left(r(u)\right) r^2(u) ds^2(\EE^{D{-}p{-}2})~.
\end{equation}
It remains now to put this metric in Hpp-wave form. For this we
write $ds^2(\EE^{p})=\sum_a d\tilde x^a d\tilde x^a$ and
$ds^2(\EE^{D{-}p{-}2})=\sum_i d\tilde y^i d\tilde y^i$. Then we
perform the following coordinate transformations
\begin{equation}
  \begin{aligned}
    u&=x^-\\
    v&=x^+ + \half \frac{\partial_-A(x^-)}{A(x^-)} x^2 + \half
    \frac{\partial_-\big(r(x^-) B(x^-)\big)}{r(x^-) B(x^-)}y^2 \\
    \tilde x^a&=\frac{1}{A(x^-)} x^a\\
    \tilde y^i&=\frac{1}{r(x^-) B(x^-)} y^i~,
  \end{aligned}
\end{equation}
where $\partial_-=\frac{d}{dx^-}$, $x^2=\delta_{ab}x^a x^b$ and
$y^2=\delta_{ij} y^i y^j$.  The metric in the new coordinate system is
\begin{multline}
  \label{plmet}
  ds^2 = 2 dx^+ dx^- + \left[ \frac{\partial_-^2 A(x^-)}{A(x^-)} x^2 +
    \frac{\partial_-^2\left(r(x^-) B(x^-)\right)}{r(x^-) B(x^-)
      }y^2\right] (dx^-)^2\\
  + ds^2(\EE^{p}) + ds^2(\EE^{D{-}p{-}2})~.
\end{multline}

It is sometimes complicated to express explicitly the non-trivial
component of the metric in the Penrose limit in terms of the $x^-$
coordinate. This is because it is difficult to find the explicit
expression for the transformation $r=r(x^-)$. However, it is
straightforward to express it in terms of the original $r$ coordinate
as follows.  Define
\begin{equation}
  \eA(x^-, x, y)=\frac{\partial_-^2 A(x^-)}{A(x^-)} x^2 +
  \frac{\partial_-^2\left(r(x^-) B(x^-)\right)}{r(x^-) B(x^-)
    }y^2
\end{equation}
Then we can use the chain rule and write
\begin{equation}
\partial_-f(x^-)=\partial_- r \partial_r f(r)=\pm \frac{1}{AB}
\partial_r f(r)
\end{equation}
where we have use that in the Penrose limit $\partial_r x^-=\pm
AB$ and $f(x^-)=f(r(x^-))$. In particular we find that
\begin{multline}\label{pla}
    \eA(r,x,y) = \left[\frac{\partial^2_r A}{A^3 B^2} -
      \frac{(\partial_r A)^2}{A^4 B^2} - \frac{\partial_r A
        \partial_r B}{A^3 B^3}\right]x^2\\
    +\left[\frac{\partial^2_r B}{A^2 B^3} + \frac{\partial_r B}{r A^2
        B^3} - \frac{\partial_r A}{r A^3 B^2} - \frac{(\partial_r
        B)^2}{A^2 B^4} - \frac{\partial_r A\partial_r B}{A^3
        B^3}\right] y^2~.
\end{multline}

Turning to investigate the Penrose limit of the form-field strengths,
we remark that they vanish at the limit. To see this observe that in
the $(u,v,\tilde x^a, \tilde y^i)$ coordinate system
\begin{equation}
  F_{p+2}=2 (-1)^p du\wedge dv\wedge \dvol(\EE^p) (\partial_u C)(r(u+v))
\end{equation}
An appropriate choice for a gauge potential is
\begin{equation}
  C_{p+1}= 2 (-1)^p dv\wedge \dvol(\EE^p) C(r(u+v))\ .
\end{equation}
The Penrose limit of the above $C_{p+1}$ is zero since, as we had seen in
\eqref{eq:PLA}, $i(\d/\d v)\bar{C}_{p+1}=0$. Thus
$F_{p+2}=dC_{p+1}=0$ in the Penrose limit. For the scalar $\phi$
in the Penrose limit we find $\phi=\phi(r(x^-))$. We remark that if a solution
has non-vanishing scalars, then it is possible to choose different frames to
describe the metric \eqref{met}. However the formulae that we have presented
above for the description of the
Penrose limit do not depend on the choice of frame. So they can easily be adapted to
any choice of frame.

Next we shall investigate the D-branes, NS-branes and M-branes
separately.

\subsubsection{D-branes}

The spacetime metric of a D$p$-brane in the string frame
\cite{GibbonsMaeda, DuffLuD3, DuffLu5Brane, HorowitzStrominger} is
\begin{equation}
\begin{aligned}[m]
  ds^2&=H^{-\frac{1}{2}} ds^2(\EE^{1,p})+H^{\frac{1}{2}}
  ds^2(\EE^{9-p})\\
F_{p+2}&=\dvol(\EE^{1,p})\wedge dH^{-1}\\
e^{2\phi}&= H^{\frac{3-p}{2}}
  \end{aligned}
\end{equation}
where
\begin{equation}
  H=1+{\frac{Q_p}{ r^{7-p}}}
\end{equation}
is a harmonic function on the transverse space $\EE^{9-p}$, $r$
is the radial coordinate in $\EE^{9-p}$ the and $Q_p$ is the
charge of D$p$-brane in some units. Note that  the form field
strength associated of the D3-brane is self-dual and so one has to
project onto the self-dual component of the field strength
presented above.

We shall focus on the Penrose limit of the Dp-brane metric. Then
at the end we shall give all non-vanishing fields of the solution.
Since $A=H^{-{\frac{1}{4}}}$ and $B=H^{{\frac{1}{4}}}$, then
$u+v=\pm r$; the integration constant has been absorbed in the
definition of the $(u,v)$ coordinates. Writing the D$p$-brane
metric in these coordinates, we have
\begin{multline}
    ds^2=2dudv+\big(2-H^{-\frac12}(u+v)\big) du^2+H^{-\frac12}(u+v) ds^2(\EE^{p})\\
    +(u+v)^2 H^{\frac12}(u+v) ds^2(S^{8-p})
\end{multline}
In the Penrose limit we get
\begin{equation}
  ds^2= 2 du dv + H(u)^{-\frac12}
  ds^2(\EE^{p}) + u^2 H(u)^{\frac12}
  ds^2(\EE^{8-p})~.
\end{equation}
It remains to put this metric in Brinkman form. For this we change
coordinates again as
\begin{equation}
  \begin{aligned}[m]
    u&=x^-\\
    v&=x^+ - \tfrac18 H^{-1}(x^-) H'(x^-) x^2 + \left( \tfrac18
       H^{-1}(x^-) H'(x^-)+ \half  (x^-)^{-1}\right) y^2 \\
    \tilde x^a&=H^{\frac{1}{4}}(x^-) x^a\\
    \tilde y^i&=(x^-)^{-1}H^{-{\frac{1}{4}}}(x^-) y^i\\
  \end{aligned}
\end{equation}
where $ds^2(\EE^p)=\sum_ad\tilde x^ad\tilde x^a$ and
$ds^2(\EE^{8-p})=\sum_id\tilde y^id\tilde y^i$ as in the general case.
Performing this coordinate transformation, we get
\begin{equation}
\begin{aligned}
    ds^2 &= 2 dx^- dx^+ + \eA(x^-,x,y) (dx^-)^2 + ds^2(\EE^{8})\\
e^{2\phi}&=H^{\frac{3-p}{2}}(x^-)
\end{aligned}
\end{equation}
where
\begin{multline}
  \eA(x^-,x,y) = \left(- \tfrac14 H^{-1} H'' + \tfrac5{16} H^{-2}
  (H')^2\right) x^2 \\
  + \left( \tfrac14  H^{-1} H'' -  \tfrac3{16} H^{-2} (H')^2 +
  \half (x^-)^{-1} H^{-1} H'\right) y^2
\end{multline}
and we have added the expression for the dilaton at the limit for
completeness.

For example let us consider the D3-brane separately. For a D3-brane,
$H=1+{\frac{Q_3}{ r^4}}$. Then $A$ can be easily computed and yields
\begin{equation}
  \eA(x^-, x,y)= -5 \frac{Q_3(x^-)^2}{((x^-)^4+Q_3)^2} x^2+3
  \frac{Q_3(x^-)^2}{((x^-)^4+Q_3)^2} y^2~.
\end{equation}

Observe that the D3-brane metric is Ricci flat in the Penrose limit and
so solves the Einstein equations without active form-field strengths
or scalars.  In the near horizon limit, $H=Q_3/r^4$, $A=0$ and the
Penrose limit is the ten-dimensional Minkowski spacetime.

The Penrose limit for Dp-branes was investigated in the string frame.
The above analysis can easily been done in other frames like for
example the Einstein frame.  In fact in such a case the result cannot
be presented in a closed form for $p\not=3$ because it is not
straightforward to give the coordinate transformation $u=u(r)$ in a
closed form. However the general formulae given in the previous
section, in particular \eqref{plmet} with \eqref{pla}, can be used to
find the Penrose limit metric.

\subsubsection{Fundamental strings and NS5-branes}

The  NS5-brane solution \cite{NS5Brane} is
\begin{equation}
  \begin{aligned}[m]
    ds^2 &= ds^2(\EE^{1,5}) + H(r) ds^2(\EE^4)\\
    F_{7}&=\dvol(\EE^{1,5})\wedge dH^{-1}\\
    e^{2\phi}&= H
  \end{aligned}
\end{equation}
where $H=1+\frac{Q_5}{r^2}$ is a harmonic function in $\EE^4$ and
$Q_5$ is the charge of NS5-brane. So in this case
$(a+b)'=\pm H^{\frac12}$.  Unfortunately, it is not possible to find
the transformation $r=r(u+v)$ explicitly. However we can still express
the non-trivial component of the metric in the Penrose limit in terms
of the $r$ coordinate as it has been explained.  Indeed we find that
the Penrose limit of the NS5-brane is
\begin{equation}
  \begin{aligned}[m]
    ds^2&=2 dx^+ dx^-+\eA(r, x,y) (dx^-)^2+ ds^2(\EE^5)+ds^2(\EE^3) \\
    e^{2\phi}&= H
  \end{aligned}
\end{equation}
where
\begin{equation}
  \eA(r,x,y)= \left[ \half H^{-2} \partial_r^2 H+ \frac{1}{2r} H^{-2}
    \partial_r H - \half H^{-3} (\partial_r H)^2\right] y^2~.
\end{equation}

In the near horizon case, we have
\begin{equation}
  (a+b)'=\pm |Q_5|^{\frac12} r^{-1}
\end{equation}
and so $r=\exp \left(\pm (u+v)/|Q_5|^{\frac12}\right)$. Choosing the
plus sign and substituting this into the metric and dilaton and taking
the Penrose limit, we find that
\begin{equation}
\begin{aligned}[m]
ds^2&=2dx^+ dx^-+
ds^2(\EE^8)\\
e^{2\phi}&= Q_5 e^{-2x^-/|Q_5|^{\frac12}}~.
\end{aligned}
\end{equation}
This is flat space with linear dilaton solution of type II (or
heterotic) supergravity and preserves sixteen (or eight)
supersymmetries.

The  fundamental string solution \cite{FString} is
\begin{equation}
  \begin{aligned}[m]
    ds^2&=H^{-1} ds^2(\EE^{1,1})+ds^2(\EE^8)\\
    F_{3}&=\dvol(\EE^{1,1})\wedge dH^{-1}\\
    e^{2\phi}&=H^{-1}
  \end{aligned}
\end{equation}
where $(a+b)'=H^{-{\frac{1}{2}}}$. As in the previous case, the
coordinate transformation $r=r(u+v)$ cannot be found explicitly.
Nevertheless we can write the Penrose limit as
\begin{equation}
  \begin{aligned}[m]
    ds^2&=2 dx^+ dx^-+\eA(r, x,y) (dx^-)^2+ ds^2(\EE^5)+ds^2(\EE^3) \\
    e^{2\phi}&= H^{-1}
  \end{aligned}
\end{equation}
where
\begin{equation}
    \eA(r,x,y)= \left[ \half H^{-1} (\partial_r H)^2- \half
      \partial^2_rH\right] x^2 + \frac{1}{2r} \partial_r H y^2~.
\end{equation}

For the near horizon case, we have $(a+b)'=\pm r^3/ |Q_1|^{\frac12}$,
so that $r^2=2 |Q_1|^{\frac14} (u+v)^{\frac12}$; we have chosen the
plus sign. Taking the Penrose limit, the metric becomes
\begin{equation}
  ds^2=2dudv +8 |Q_1|^{-{\frac{1}{4}}} u^{{\frac{3}{2}}} ds^2(\EE)
  +2 |Q_1|^{{\frac{1}{4}}} u^{\frac12} ds^2(\EE^7)
\end{equation}
In Brinkman coordinates, the Penrose limit of
the fundamental string solution in the near horizon limit is
\begin{equation}
\begin{aligned}[m]
  ds^2 &= 2 dx^- dx^+ + \eA(x^-,x,y) (dx^-)^2 + ds^2(\EE) + ds^2(\EE^7)
  \\
  e^{2\phi}&=8 |Q_1|^{-{\frac{1}{4}}} (x^-)^{{\frac{3}{2}}}~,
  \end{aligned}
\end{equation}
where
\begin{equation}
 \label{eq:PLfsA}
  \eA(x^-,x,y) = - \tfrac3{16} (x^-)^{-2} x^2 - \tfrac3{16} (x^-)^{-2} y^2\ .
\end{equation}
The metric is actually Lorentzian homogeneous (cf.\ the discussion in
Section~\ref{sec:frw}). Namely, in addition to the $2D{-}3{=}17$
Heisenberg algebra Killing vectors \eqref{eq:heisenberg} of a generic
pp-wave spacetime, there is the $SO(8)$ rotation symmetry of the
transverse coordinates, and there is the scale invariance
$(x^+,x^-)\to (cx^+,c^{-1} x^-)$ corresponding to the Killing vector
$x^+\d_+ - x^-\d_-$. Since the dilaton depends non-trivially on $x^-$,
however, only the 45-dimensional subgroup of the isometry group
generated by the Heisenberg algebra and the transverse $SO(8)$ is a
symmetry of the solution. This group does not act transitively on the
spacetime.

\subsubsection{M-branes}

There are two cases to consider, the M2-brane and the M5-brane.
The supergravity solution for the M2-brane \cite{DS2brane} is
\begin{equation}
  \begin{aligned}[m]
    ds^2&=H^{-\frac23} ds^2(\EE^{1,2}) + H^{\frac13} ds^2(\EE^8)\\
    F_4&=\dvol(\EE^{1,3})\wedge dH^{-1}
  \end{aligned}
\end{equation}
Thus we have
\begin{equation}
  (a+b)'=H^{-\frac16}~.
\end{equation}
This differential equation cannot be easily integrated.
Nevertheless we can write the Penrose limit metric as
\begin{equation}
  ds^2=2 dx^+ dx^-+\eA(r, x,y) (dx^-)^2+ ds^2(\EE^2)+ds^2(\EE^7)
\end{equation}
where
\begin{multline}
    \eA(r, x, y)=\left[ \tfrac7{18} H^{-\frac53} (\partial_r H)^2 -
      \tfrac13 H^{-\frac23} \partial_r^2 H\right] x^2\\
    + \left[\tfrac16 H^{-\frac23} \partial_r^2 H  + \tfrac1{2r} H^{-\frac23}
      \partial_r H - \tfrac19 H^{-\frac53} (\partial_r H)^2\right]
    y^2~.
\end{multline}
In the near horizon limit, we can easily find that
\begin{equation}
  (u+v)=\pm \half  |Q|^{-\frac16} r^2
\end{equation}
The near horizon M2-brane metric becomes
\begin{multline}
  ds^2 = 2 du dv + \left(2-4 |Q_2|^{-\frac13} (v+u)^2\right) du^2\\
  + 4 |Q_2|^{-\frac13} (v+u)^2 ds^2(\EE^2) + |Q_2|^{\frac13}
  ds^2(S^7)~.
\end{multline}
The Penrose limit gives
\begin{equation}
  ds^2=2dudv+ 4 |Q_2|^{-\frac13} u^2 ds^2(\EE^2) + |Q_2|^{\frac13}
  ds^2(\EE^7)~,
\end{equation}
This in fact is the eleven-dimensional Minkowski spacetime as it can
be easily seen by writing the metric in Brinkman coordinates. This is
in agreement with the general result obtained in
Section~\ref{sec:PLadss1} that the Penrose limit of any $\AdS\times S$
space along a radial geodesic is flat.

Turning now to investigate the M5-brane. The supergravity solution
for the M5-brane \cite{Guven} is
\begin{equation}
  \begin{aligned}[m]
    ds^2&=H^{-\frac13} ds^2(\EE^{1,5}) + H^{\frac23} ds^2(\EE^5)\\
    F_7&=\dvol(\EE^{1,5})\wedge dH^{-1}
  \end{aligned}
\end{equation}
Thus we have that $(a+b)'=H^{\frac16}$. This differential
equation cannot be easily integrated. Nevertheless we can write
the Penrose limit metric as
\begin{equation}
  ds^2=2 dx^+ dx^-+\eA(r, x,y) (dx^-)^2+ ds^2(\EE^5)+ds^2(\EE^4)
\end{equation}
where
\begin{multline}
   \eA(r, x, y) = \left[-\tfrac16 H^{-\frac43} \partial_r^2H + \tfrac29
      H^{-\frac73} (\partial_r H)^2\right] x^2\\
    +\left[\tfrac13 H^{-\tfrac43} \partial_r^2 H +
    \tfrac1{2r} H^{-\frac43} \partial_r H - \tfrac5{18} H^{-\frac73}
    (\partial_r H)^2\right] y^2~.
\end{multline}

Again in the near horizon limit we can find
\begin{equation}
  (u+v)^2=4|Q_5|^{\frac13} r
\end{equation}
Substituting this back into near horizon geometry and taking the
Penrose limit we find
\begin{equation}
  ds^2=2dudv+ |Q_5|^{-\frac23} u^2 ds^2(\EE^5) + |Q_5|^{\frac23}
  ds^2(\EE^4)
\end{equation}
Noting that the non-trivial coefficients of the metric are at most
quadratic in $u$ and using \eqref{eq:bi}, or changing coordinates as
for the M2-brane above, we see that this metric is in fact, as
expected from our general arguments, eleven-dimensional Minkowski
spacetime.

\subsection{Generic null geodesics}

We now consider the Penrose limit of brane solutions along generic
null geodesics; that is, geodesics whose tangent vectors have a
component tangent to the transverse sphere.

Symmetry considerations allow us to single out any direction on the
transverse sphere.  To this end we will write the sphere metric as
\begin{equation}
  d\Omega_{D-p-2}^2 = d\psi^2 + (\sin\psi)^2 d\Omega_{D-p-3}^2~,
\end{equation}
where $\psi$ is a colatitude and $d\Omega_{D-p-3}^2$ is the metric on
the corresponding equator.  The brane metric becomes
\begin{equation}
  ds^2 = A^2 \left(-dt^2 + ds^2(\EE^p)\right) + B^2 dr^2 + B^2 r^2
  \left( d\psi^2 +  (\sin\psi)^2 d\Omega_{D-p-3}^2\right)~.
\end{equation}
We will consider null geodesics in the $(t,r,\psi)$ space with metric
\begin{equation}
  ds^2 = - A^2 dt^2 + B^2 dr^2 + B^2 r^2 d\psi^2~.
\end{equation}
Let us change coordinates to $(u,v,z)$ adapted to the null geodesic:
\begin{equation}
  u = u(r) \qquad v = t + \ell \psi + a(r) \qquad z = \psi + b(r)~,
\end{equation}
where $\ell$ is a constant, and such that the metric takes the form
\begin{equation}
  ds^2 = 2 du dv + K dv^2 + L dv dz + M dz^2~.
\end{equation}
This choice of reparametrisation is consistent with a null geodesic
with tangent vector
\begin{equation}
  \frac{\d}{\d u} = f(r) \frac{\d}{\d r} + g(r) \frac{\d}{\d \psi} +
  h(r) \frac{\d}{\d t}
\end{equation}
which has $r$-dependent components in all three directions.  The
constant parameter $\ell$ can be understood as the angular momentum of
the massless particle whose motion is described by the null geodesic.
Comparing the forms of the metric in both coordinate systems we can
determine $K$, $L$ and $M$, the functions $f$,$g$ and $h$, the
function $u(r)$ and the functions $a$, $b$ in terms of the parameter
$\ell$ and the functions $A$ and $B$ appearing in the metric.

After some calculation, and letting primes denote differentiation with
respect to $r$, we find the following
\begin{gather}
  f^2 = \frac{1}{B^2} \left(\frac{1}{A^2} -
    \frac{\ell^2}{B^2r^2}\right) \qquad
  h = -\frac{1}{A^2} \qquad g = \frac{\ell}{B^2 r^2}\\
  K = - A^2 \qquad L = 2\ell A^2 \qquad M = B^2r^2 - \ell A^2\\
  a'= \pm \sqrt{\frac{B^2}{A^2} - \frac{\ell^2}{r^2}} \qquad
  b' = \mp \frac{\ell/r^2}{\sqrt{\frac{B^2}{A^2} -
      \frac{\ell^2}{r^2}}}~,
\end{gather}
and
\begin{equation}
 \label{eq:Q}
 \frac{du}{dr}= \pm\frac{B^2 }{\sqrt{\frac{B^2}{A^2} -
     \frac{\ell^2}{r^2}}}= \pm \frac{r A B^2  }{\sqrt{r^2 B^2-\ell^2
     A^2}} \equiv Q
\end{equation}
So $u(r)$ is defined up to an inconsequential constant of integration
by the following integral
\begin{equation}
  u(r) = \pm \int^r \frac{B^2 dr}{\sqrt{\frac{B^2}{A^2} -
  \frac{\ell^2}{r^2}}}~.
\end{equation}
This equation can be inverted to give an implicit relation $r(u)$
which will play an important role below.  All the above signs are
correlated and come from choosing the sign of the square root of
$f^2$.

In the Penrose limit we obtain the following spacetime metric in Rosen
coordinates:
\begin{multline}
  \label{eq:genericrosen}
  ds^2 = 2 du dv + \left(B^2 r^2 - \ell^2 A^2 \right) dz^2 + A^2
  ds^2(\EE^p)\\   + B^2 r^2 (\sin b)^2 ds^2(\EE^{D{-}p{-}3})~,
\end{multline}
where the dependence on $u$ is implicit through the dependence on $r$.
In the limit $\ell \to 0$, we recover the result for the Penrose
limits associated with the radial null geodesics investigated in the
previous sections provided that we choose $b=\pi/2$.

The field strength for a $p$-brane solution is of the form
\begin{equation}
  F_{p+2} = \dvol(\EE^{1,p}) \wedge dC(r)~,
\end{equation}
for some function $C(r)$.  Changing coordinates and taking the Penrose
limit, we find the following field strength
\begin{equation}
  \bar F_{p+2} = \pm C' \frac{\ell}{B} \sqrt{\frac{1}{A^2} -
  \frac{\ell^2}{B^2r^2}} du \wedge dy^1 \wedge \dots \wedge dy^p
  \wedge dz~,
\end{equation}
which is nonzero provided that $\ell$ is different from zero.

It remains to write the metric in pp-wave form. For this we write
$ds^2(\EE^{p})=\sum_a d\tilde x^a d\tilde x^a$,
$ds^2(\EE^{D{-}p{-}3})=\sum_i d\tilde y^i d\tilde y^i$ and $\tilde
z=z$. Then we perform the following coordinate transformations
\begin{equation}
  \begin{aligned}
    u&=x^-\\
    v&=x^+ + \half\frac{\partial_-A(x^-)}{A(x^-)} x^2 + \half
    \frac{\partial_-\big(r(x^-) B(x^-)\sin b\big)}{r(x^-) B(x^-)\sin b}y^2
    \\
    &~~~~~~~~~~~~~~+\half
   \frac{ \partial_- (\sqrt{B^2 r^2-\ell^2A^2})}{\sqrt{B^2 r^2-\ell^2A^2}}z^2 \\
    \tilde x^a&=\frac{1}{A(x^-)} x^a\\
    \tilde y^i&=\frac{1}{r(x^-) B(x^-)\sin b} y^i\\
    \tilde z&=\frac{1}{\sqrt{B^2 r^2-\ell^2A^2}} z~,
  \end{aligned}
\end{equation}
where $\partial_-=\frac{d}{dx^-}$.  The metric in the new coordinate
system is
\begin{equation}
\label{plmet2}
 ds^2=2 dx^+ dx^- + \eA(x^-,x,y,z)(dx^-)^2
  +ds^2(\EE^p) + ds^2(\EE^{D-p-3})+dz^2~,
 \end{equation}
where
\begin{multline}
    \eA(x^-, x, y,z)= \frac{\partial_-^2 A(x^-)}{A(x^-)} x^2 +
    \frac{\partial_-^2\left(r(x^-) B(x^-)\sin b \right)}{r(x^-)
      B(x^-)\sin b}y^2 \\
    +\frac{\partial_-^2\sqrt{B^2 r^2-\ell^2A^2}}{\sqrt{B^2
        r^2-\ell^2A^2}} z^2~.
\end{multline}

As in \eqref{pla} we can express the non-trivial component of the
above metric as a function of the original coordinate $r$ using the
formula
\begin{equation}
\partial_-^2 f(x^-)=- Q^{-3} \partial_rQ \partial_r f(r)+ Q^{-2} \partial_r^2 f(r)~,
\end{equation}
where $Q$ was defined in \eqref{eq:Q}.  The resulting metric is
\begin{equation}
ds^2= 2 dx^+ dx^- +\eA(r, x,y,z) (dx^-)^2+ds^2(\EE^{D-2})
\end{equation}
where
\begin{multline}
  \label{plgn}
  \eA(r,x,y,z)= (B^2 r^2-\ell^2A^2)^{-\half}
  \big[-Q^{-3}\partial_r Q \partial
  \sqrt{B^2r^2-\ell^2 A^2}\\ +Q^{-2} \partial_r^2\sqrt{B^2r^2-\ell^2
    A^2}\big] z^2 + A^{-1} \big[ -Q^{-3} \partial_rQ \partial_r A+
  Q^{-2}\partial_r^2 A\big] x^2\\
   +(Br\sin b)^{-1} \big[ -Q^{-3}\partial_r Q \partial_r(Br \sin
  b)+Q^{-2} \partial^2_r(Br \sin b)\big] y^2
\end{multline}

Unfortunately, there does not seem to be a straightforward way to
simplify this equation. So we shall not investigate each brane case
separately as we have done for the case of radial null geodesics. We
remark though that in the limit that $\ell\to 0$, the above
calculations and formulae reduce to those we have found for the radial
null geodesics. In addition in the near horizon limit for the M-branes
and D3-brane, one can show after some computation that the above
calculations reduce to those we have done for the Penrose limit along
a generic null geodesic of the associated $\AdS\times S$ spaces. The
near horizon geometries of the various branes have been investigated
in \cite{GibbonsTownsend}.

\section{Penrose limits of intersecting branes}
\label{sec:intersections}

The methods developed to investigate the Penrose limits of branes can
be easily adapted to investigate the Penrose limit of intersecting
brane solutions
\cite{PTIntersections,TseytlinHarmonic,GKTOverlapping}. Since the
solutions for intersecting branes localised at the same point in the
overall transverse space are of cohomogeneity one, the same symmetry
considerations that have been employed for branes apply to classify
all possible directions of null geodesics.  There are three types
depending on the direction $V$ of the null geodesic at a point. The
tangent bundle of an intersecting brane solution can split as
$T_PM=T_PB\oplus T_PR\oplus T_PS$, where $T_PB$ spans the common
intersection and relative transverse directions, $T_PR$ spans the
direction along the radial coordinate and $T_PS$ spans the directions
along the overall transverse sphere. Thus we have to consider the
cases of (i) longitudinal null geodesics, (ii) radial null geodesics
and (iii) generic null geodesics whose definitions are parallel to
those in the case of branes.  The only subtlety is that the isotropy
group of the point $P$ does not act transitively on the timelike
directions in $T_PB$.  This complicates the classification of Penrose
limits, as we have to distinguish between different classes of generic
null geodesics, depending on whether the velocity vector has a
component (or not) along each of relative transverse directions.  We
will only consider Penrose limits along those generic null geodesics
whose velocity component in $TB$ is tangent to the common
intersection.  The general case is straightforward.

It can be easily seen that the Penrose limit of an intersecting brane
solution along longitudinal null geodesics is Minkowski spacetime; the
form-field strengths vanish and the scalars are constant. So it
remains to investigate Penrose limits along the radial null geodesics
and the generic null geodesics.  There is a large number of
intersecting brane configurations. Therefore instead of doing the
analysis for each case separately, we shall focus on general formulae
adapted to the types of solution associated with intersecting branes.
Then we give some representative examples associated with
supersymmetric intersecting M-branes.

\subsection{General formulae for intersecting branes}
We begin with some general formulae for the Penrose limits of
intersecting branes along radial and generic null geodesics.

\subsubsection{Radial null geodesics}

The metric of a typical intersecting brane solution is
\begin{equation}
ds^2=A^2 ds^2(\EE^{1,p})+\sum^k_{i=1} A_i^2 ds^2_i(\EE^{n_i})
+ B^2 ds^2(\EE^{D-p-n-1})
\end{equation}
where $\EE^{1,p}$ are the coordinates along the common intersection,
$\EE^{n_i}$ are the relative transverse spaces, $\EE^{D-p-n-1}$ is the
overall transverse space and $n=\sum_i n_i$.  We shall consider
solutions for which all components $A^2, A_i^2, B^2$ depend on the
radial coordinate $r$ of the overall transverse space $B^2$.  To
continue we rewrite the metric as
\begin{equation}
ds^2=A^2 ds^2(\EE^{1,p})+\sum^k_{i=1} A_i^2 ds^2_i(\EE^{n_i})
+ B^2(dr^2+r^2 ds^2(S^{D-p-n-2})) \ .
\end{equation}
The investigate the Penrose limit for radial null geodesics for
intersecting branes is similar to that for the standard brane
solutions investigated in the previous sections.  The relevant part of
the metric is again the two-dimensional metric
\begin{equation}
ds^2_{(2)}=-A^2 dt^2 +B^2 dr^2 \ .
\end{equation}
After performing a coordinate transformation similar to that in the
case of branes and taking the Penrose limit, we find that the metric
becomes
\begin{equation}
ds^2=2 du dv+A^2 ds^2(\EE^{p})+\sum^k_{i=1} A_i^2 ds^2_i(\EE^{n_i})
+ B^2 r^2 ds^2(\EE^{D-p-n-2}) \,
\end{equation}
where the components of the metric depend on $u$, and $u$ is related
to the radial coordinate $r$ by
\begin{equation}
  \frac{du}{dr}= A^2 B^2
\end{equation}
To put the metric in the standard pp-wave form, we write
$ds^2(\EE^p)=\sum_a (d\tilde x^a)^2$, $ds^2(\EE^{n_i})=\sum_{a_i}
(d\tilde x^{a_i}_i)^2$ and $ds^2=\sum_{p} (d\tilde y^p)^2$ and perform
the coordinate transformation
\begin{equation}
  \begin{aligned}
    u&=x^-\\
    v&=x^+ + \half \frac{\partial_-A(x^-)}{A(x^-)} x^2 +\half \sum_i
     \frac{\partial_-A_i(x^-)}{A_i(x^-)} x_i^2+ \half
    \frac{\partial_-\big(r(x^-) B(x^-)\big)}{r(x^-) B(x^-)} \\
    \tilde x^a&=\frac{1}{A(x^-)} x^a\\
    \tilde x_i^{a_i}&=\frac{1}{A_i(x^-)} x_i^{a_i}\\
    \tilde y^p&=\frac{1}{r(x^-) B(x^-)} y^p~,
  \end{aligned}
\end{equation}
where $\partial_-=\frac{d}{dx^-}$.  The metric in the new coordinate 
system is
\begin{equation}
  \label{plmet3}
  ds^2=2 dx^+ dx^- +\eA(x^-,x,x_i,y)  (dx^-)^2
  + ds^2(\EE^{D-2}) ~,
\end{equation}
where
\begin{equation}
\eA(x^-,x,x_i,y)= \frac{\partial_-^2 A(x^-)}{A(x^-)} x^2 +
  \sum_i\frac{\partial_-^2 A_i(x^-)}{A_i(x^-)} x_i^2+
    \frac{\partial_-^2\left(r(x^-) B(x^-)\right)}{r(x^-) B(x^-)
      }y^2~.
\end{equation}

Again, it is sometimes complicated to express explicitly the
non-trivial component of the metric in the Penrose limit in terms of
the $x^-$ coordinate. This is because it is difficult to find the
explicit expression for the transformation $r=r(x^-)$. However using
the chain rule, it is straightforward to express $\eA$ in terms of the
original $r$ coordinate as follows:
\begin{multline}\label{plib}
    \eA(r,x, x_i,y) = \left[\frac{\partial^2_r A}{A^3 B^2} -
      \frac{(\partial_r A)^2}{A^4 B^2} - \frac{\partial_r A
        \partial_r B}{A^3 B^3}\right]x^2\\
        +\sum_i\left[\frac{\partial^2_r A_i}{A^2 B^2 A_i} -
      \frac{\partial_r (AB) \partial_r A_i}{A^3 B^3 A_i}\right] x_i^2\\
       +\left[\frac{\partial^2_r B}{A^2 B^3} + \frac{\partial_r B}{r A^2
        B^3} - \frac{\partial_r A}{r A^3 B^2} - \frac{(\partial_r
        B)^2}{A^2 B^4} - \frac{\partial_r A\partial_r B}{A^3
        B^3}\right] y^2~.
\end{multline}

\subsubsection{Generic null geodesics}

We now consider the Penrose limit of intersecting brane solutions
along generic null geodesics; that is, geodesics whose tangent vectors
have a component tangent to the overall transverse sphere. The
investigation of the Penrose limit along such geodesics is similar to
that explained for brane solutions. So we shall not elaborate.  As
mentioned above, we will only consider null geodesics whose velocity
component in $TB$ is tangent to the common intersection.

First we write the metric of the intersecting brane solution as follows:
\begin{equation}
\begin{aligned}[m]
  ds^2 = A^2 \left(-dt^2 + ds^2(\EE^p)\right)&+\sum_i A^2_i(r)
  ds_i^2(\EE^{n_i}) + B^2 dr^2 \\
  &+ B^2 r^2
  \left( d\psi^2 +  (\sin\psi)^2 d\Omega_{D-p-3}^2\right)~.
  \end{aligned}
\end{equation}
Then we consider null geodesics in the $(t,r,\psi)$ space with induced
metric
\begin{equation}
  ds^2 = - A^2 dt^2 + B^2 dr^2 + B^2 r^2 d\psi^2~.
\end{equation}
and change coordinates to $(u,v,z)$ adapted to the null geodesic as
\begin{equation}
  u = u(r) \qquad v = t + \ell \psi + a(r) \qquad z = \psi + b(r)~,
\end{equation}
where $\ell$ is a constant, and such that the metric takes the form
\begin{equation}
  ds^2 = 2 du dv + K dv^2 + L dv dz + M dz^2~.
\end{equation}
Since the relevant three dimensional metric is the same as that of
Penrose limits for branes, the various formulae for the coordinate
transformations are the same. The final result is
\begin{equation}
\label{plintmet}
  ds^2= 2dx^+ dx^- + \eA(x^-, x, y,z)(dx^-)^2
  + ds^2(\EE^{p}) + ds^2(\EE^{D{-}p{-}3})+dz^2~,
\end{equation}
where and
\begin{equation}
\begin{aligned}[m]
\eA(x^-, x, x_i, y,z)= &\frac{\partial_-^2 A(x^-)}{A(x^-)} x^2
+ \sum_i \frac{\partial_-^2 A_i(x^-)}{A_i(x^-)}
x_i^2
\\ &
+ \frac{\partial_-^2\left(r(x^-) B(x^-)\sin b \right)}{r(x^-) B(x^-)\sin b}y^2
+\frac{\partial_-^2\sqrt{B^2 r^2-\ell^2A^2}}{\sqrt{B^2 r^2-\ell^2A^2}} z^2~.
\end{aligned}
\end{equation}

As in \eqref{pla} we can express the non-trivial component of the
above metric as a function of the original coordinate $r$ using the
formula
\begin{equation}
\partial_-^2 f(x^-)=- Q^{-3} \partial_rQ \partial_r f(r)+ Q^{-2} \partial_r^2 f(r)~,
\end{equation}
where $Q$ was defined in \eqref{eq:Q}.  The resulting metric is
\begin{equation}
ds^2= 2 dx^+ dx^- +\eA(r, x,y,z) (dx^-)^2+ds^2(\EE^{D-2})
\end{equation}
where
\begin{equation}
\begin{aligned}\label{plintgn}
  \eA(r,x,y,z)&= (B^2 r^2-\ell^2A^2)^{-\half}
  \big[-Q^{-3}\partial_r Q \partial
  \sqrt{B^2r^2-\ell^2 A^2}\\ &+Q^{-2} \partial_r^2\sqrt{B^2r^2-\ell^2 A^2}\big] z^2\\
  &+ A^{-1} \big[ -Q^{-3} \partial_rQ \partial_r A+ Q^{-2}\partial_r^2 A\big] x^2\\
  &+\sum_i A_i^{-1} \big[ -Q^{-3} \partial_rQ \partial_r A_i+ Q^{-2}\partial_r^2 A_i\big] x_i^2\\
  & +(Br\sin b)^{-1} \big[ -Q^{-3}\partial_r Q \partial_r(Br \sin
  b)+Q^{-2} \partial^2_r(Br \sin b)\big] y^2
\end{aligned}
\end{equation}

As in the case of branes, there does not seem to be a straightforward
way to simplify this equation.  We remark though that in the limit
that $\ell\to 0$, the above calculations and formulae reduce to those
we have found for the radial null geodesics. In addition for those
intersecting brane configurations for which the near horizon limit is
$\AdS\times S\times \EE$, the Penrose limit along a generic geodesics
is $P\times \EE$ where $P$ is the Penrose limit of the associated
$\AdS\times S$ spacetime.

\subsection{Three M2-branes intersecting on a $0$-brane}

The intersecting brane configuration that we shall consider is
\begin{equation}
  \begin{matrix}
    M2& 0 & 1 & 2 &   &   &   &\\
    M2& 0 &   &   & 3 & 4 &   &\\
    M2& 0 &   &   &   &   & 5 & 6
  \end{matrix}
\end{equation}

Associating a harmonic function with each brane involved in the
configuration, the spacetime solution is
\begin{equation}
  \label{eq:mmm}
  \begin{aligned}
    ds^2&=H_1^{\frac13} H_2^{\frac13} H_3^{\frac13}
    \big(-H_1^{-1}H_2^{-1}H_3^{-1}dt^2 + H_1^{-1} ds_1(\EE^2)\\
    & \qquad {} + H_2^{-1} ds_2(\EE^2) + H_3^{-1} ds_3(\EE^2) +
    ds^2(\EE^4)\big)\\
    F_4&=dt\wedge [\dvol_1(\EE^2)\wedge dH_1^{-1}+\dvol_2(\EE^2)\wedge
    dH_2^{-1}+ \dvol_3(\EE^2)\wedge dH_3^{-1}]
  \end{aligned}
\end{equation}
where $H_i=1+\frac{|Q_i|}{r^2}$ are harmonic functions on $\EE^4$.

In this case $(a+b)'=H_1^{-\frac16} H_2^{-\frac16}H_3^{-\frac16}$.
The transformations $r=r(u+v)$ cannot be found explicitly but the
non-trivial component of the associated pp-wave metric is given by
\eqref{plib} for
\begin{equation}
\begin{aligned}[m]
A&=H_1^{-\frac13}H_2^{-\frac13}H_3^{-\frac13}\\
A_1&=H_1^{-\frac13}H_2^{\frac16}H_3^{\frac16}\\
A_2&=H_1^{\frac16}H_2^{-\frac13}H_3^{\frac16}\\
A_3&=H_1^{\frac16}H_2^{\frac16}H_3^{-\frac13}\\
B&=H_1^{\frac16}H_2^{\frac16}H_3^{\frac16}
\end{aligned}
\end{equation}

The calculation can be performed in a closed form in the special case
where $H_1=H_2=H_3=H$.  The metric in this case is
\begin{equation}
  ds^2= -H^{-2} dt^2 + ds^2(\EE^2) + ds^2(\EE^2) + ds^2(\EE^2) + H
  ds^(\EE^4)
\end{equation}
where $H=1+ \frac{|Q|}{r^2}$. In angular coordinates this metric can
be written as
\begin{equation}
  ds^2= -H^{-2} dt^2+ds^2(\EE^2)+ds^2(\EE^2)+ds^2(\EE^2)+H (dr^2+r^2
  ds^2(S^3))
\end{equation}
To take the Penrose limit again we have to change coordinates as in
the case of branes from $(t, r)$ to $(u,v)$. This is easily done in
this case because
\begin{equation}
  \frac{d}{dr}(u+v)=\pm H^{-\frac12}
\end{equation}
and so $(u+v)^2= r^2+|Q|$. The metric in the Penrose limit is
\begin{equation}
  ds^2=2du dv+ds^2(\EE^2)+ds^2(\EE^2)+ds^2(\EE^2)+ u^2 ds^2(\EE^3)
\end{equation}
This in fact is the Minkowski metric in eleven dimension. The Penrose
limit for the near horizon geometry of \eqref{eq:mmm} is again the
Minkowski spacetime.

Now instead of taking the Penrose limit along a radial null geodesic,
we take the Penrose limit of the same configuration along a generic
null geodesic at the near horizon limit. The near horizon geometry of
this intersecting brane solution is $\AdS_2\times S^3\times \EE^6$. The
Penrose limit of this geometry is $\CW_5\times \EE^6$, where $\CW_5$
is the Cahen-Wallace space associated which is the Penrose limit of
$\AdS_2\times S^3$.

\subsection{Two M2-branes and two M5-branes intersection on a $0$-brane}

The intersecting brane configuration that we shall consider is
\begin{equation}
  \begin{matrix}
    M2& 0 & 1 & 2 &   &   &   &   &\\
    M2& 0 &   &   & 3 & 4 &   &   &\\
    M5& 0 & 1 &   & 3 &   & 5 & 6 & 7\\
    M5& 0 &   & 2 &   & 4 & 5 & 6 & 7
  \end{matrix}
\end{equation}
Associating a harmonic function with each brane involved in the
configuration, the spacetime solution is
\begin{equation}
  \begin{aligned}
    ds^2&= H_1^{\frac13} H_2^{\frac13} H_3^{\frac23}
    H_4^{\frac23}\big(-H_1^{-1} H_2^{-1} H_3^{-1} H_4^{-1} dt^2 +
    H^{-1}_1 H^{-1}_3 ds_1^2(\EE) \\
    & \qquad {} + H^{-1}_1 H^{-1}_4 ds_2^2(\EE) + H_2^{-1}H^{-1}_3
    ds_3(\EE)+H_2^{-1} H^{-1}_4 ds_4(\EE)\\
    & \qquad {} + H^{-1}_3 H^{-1}_4 ds^2(\EE^3) + ds^2(\EE^3)\big)\\
    F_4&=dt\wedge [\dvol_1(\EE^2)\wedge dH_1^{-1}+\dvol_2(\EE^2)\wedge
    dH_2^{-1}+ \dvol_3(\EE^2)\wedge dH_3^{-1}]
  \end{aligned}
\end{equation}
where $H_i=1+\frac{Q_i}{r}$ are harmonic functions on $\EE^3$ and
$\star$ is the Hodge operation in $\EE^3$.

In this case $(a+b)'= H_1^{-\frac16} H_2^{-\frac16} H_3^{\frac1{}}
H_4^{\frac1{}}$.  The transformations $r=r(u+v)$ cannot be found
explicitly as in the previous intersecting branes example. However
we can express the non-trivial component of the Penrose limit metric
using \eqref{plib}. In particular, we take
\begin{equation}
  \begin{split}
    A&=H_1^{-\frac13}H_2^{-\frac13} H_3^{-\frac16}H_4^{-\frac16}\\
    A_1&=H_1^{-\frac13}H_2^{\frac16} H_3^{-\frac16}H_4^{\frac13}\\
    A_2&=H_1^{-\frac13}H_2^{\frac16} H_3^{\frac13}H_4^{-\frac16}\\
    A_3&=H_1^{\frac16}H_2^{-\frac13} H_3^{-\frac16}H_4^{\frac13}\\
    A_4&=H_1^{\frac16}H_2^{-\frac13} H_3^{\frac13}H_4^{-\frac16}\\
    A_5&=H_1^{\frac16}H_2^{\frac16} H_3^{-\frac16}H_4^{-\frac16}\\
    B&=H_1^{\frac16}H_2^{\frac16} H_3^{\frac13}H_4^{\frac13}
  \end{split}
\end{equation}

The calculation can be performed in a closed form in the special case
where $H_1=H_2=H_3=H_4=H$. (In fact it suffices to take $H_1=H_3$ and
$H_2=H_4$.)  The metric in eleven dimensions is
\begin{equation}
  ds^2=-H^{-2} dt^2+ds^2(\EE^7)+H^{2}ds^2(\EE^3)
\end{equation}
where $H=1+\frac{|Q|}{r}$. Changing coordinates from $(t,r)$ to
$(u,v)$, we find that $u+v=\pm r$. Taking the limit, we find the
metric
\begin{equation}
  ds^2=2dudv+ds^2(\EE^7)+ u^2 (1+\frac{|Q|}{u})^2 ds^2(\EE^2)
\end{equation}
Again this is the flat Minkowski eleven-dimensional metric.  The
Penrose limit of the associated near horizon geometry is also the
Minkowski spacetime.

Now instead of taking the Penrose limit along a radial null geodesic,
we take the Penrose limit of the same configuration along a generic
null geodesic at the near horizon limit. The near horizon geometry of
this intersecting brane solution is $\AdS_2\times S^2\times \EE^7$. The
Penrose limit of this geometry is $\CW_4\times \EE^7$, where $\CW_4$
is the Cahen-Wallace space associated which is the Penrose limit of
$\AdS_2\times S^2$.

\section{Penrose limits of supersymmetric black holes and strings in
  lower dimensions}
\label{sec:blackholes}

We shall mainly focus on the Penrose limits of supersymmetric black
holes that arise in toroidal compactifications of M- and string
theories. These solutions are again of cohomogeneity one. For black holes,
there are two types of null geodesics to consider the following: (i)
radial null geodesics and (ii) generic null geodesics.  The Penrose
limit and various formulae in this case are similar to those for $p=0$
branes. In what follows we shall mainly focus on the metric of these
black holes.

For string solutions that arise in toroidal compactifications of M-
and string theories there are three different choices of null
geodesics.  These are precisely those that occur for the p-brane, $p>0$,
solutions.

\subsection{Five-dimensional black holes}

The five-dimensional black holes that arise in toroidal
compactifications of M- and string theories can be thought off as
reduction to five-dimensions of intersecting brane configurations in
ten and eleven dimensions \cite{PTIntersections, KlebTseyInt}. We
shall use the form of such four- and five-dimensional black hole
solutions as summarised in \cite{TseyBlackHoles}.  The relevant
five-dimensional black hole metric can be written in the form
\begin{equation}
  ds^2 = - (H_1 H_2 H_3)^{-\frac23} dt^2 + (H_1 H_2 H_3)^{\frac13}
  ds^2(\EE^4)~,
\end{equation}
where $H_i=1+\frac{Q_i}{r^2}$. The harmonic functions $H_i$ are
inherited from those of the intersecting brane configuration in ten or
eleven dimensions.

For the black hole metric above, the Penrose limit along a radial null
geodesic cannot be given in a closed form because the coordinate
transformation $u=u(r)$ is not known explicitly. However, the
non-trivial component of the Penrose limit metric in the pp-wave form
can expressed in terms of the radial coordinate $r$. The metric is
\begin{multline}
  ds^2= 2 dx^+ dx^-+\big[\tfrac16 (H_1 H_2
  H_3)^{-\frac23}\partial^2_r (H_1 H_2 H_3)\\
  +\frac{1}{2r}(H_1 H_2 H_3)^{-\frac23}\partial_r (H_1 H_2 H_3)\\
  - \tfrac{1}{9} (H_1 H_2 H_3)^{-\frac53}(\partial_r (H_1 H_2
  H_3))^2\big] y^2 (dx^-)^2 +ds^2(\EE^3)~.
\end{multline}

The near horizon geometry of this black hole is $\AdS_2\times S^3$.
Thus the Penrose limit in this case is (i) five-dimensional Minkowski
spacetime if it is taken along a radial null geodesic or (ii) $\CW_5$
if it is taken along a generic null geodesic.

\subsection{Four-dimensional black holes}

The relevant metric of four-dimensional black holes which arise in
toroidal M- and string theory compactifications is
\begin{equation}
  ds^2= - (H_1 H_2 H_3 H_4)^{-2} dt^2 + (H_1 H_2 H_3 H_4)^2 ds^2(\EE^3)~,
\end{equation}
where $H_i=1+\frac{Q_i}{r}$.

In this case it is easy to see that if the limit is taken along a
radial null geodesic $u=\pm r$.  So the Penrose limit metric can be
given explicitly. In particular, we find that
\begin{multline}
    ds^2=2 dx^+ dx^-+\big[(H_1 H_2 H_3 H_4)^{-1} \partial_-^2(H_1 H_2
    H_3 H_4)\\
    + (x^- H_1 H_2 H_3 H_4)^{-1} \partial_-(H_1 H_2 H_3 H_4) \big]
    y^2 (dx^-)^2 + ds^2(\EE^2)~.
\end{multline}

The near horizon geometry of this black hole is $\AdS_2\times S^2$.
Thus the Penrose limit in this case is (i) four-dimensional Minkowski
spacetime if it is taken along a radial null geodesic or (ii) $\CW_4$
if it is taken along a generic null geodesic.

\subsection{A string in six dimensions}
\label{sec:sdstring}
A string solution in six dimensions that arise in toroidal M- and
string theory compactifications is
\begin{equation}
  ds^2= (H_1 H_2)^{-\frac12} ds^2 (\EE^2) + (H_1 H_2)^{\frac12}
  ds^2(\EE^4)~,
\end{equation}
where $H_i=1+\frac{Q_i}{r^2}$. The self-dual string solution arises
for $H_1=H_2$ \cite{DuffLuSDS}.

For Penrose limits along longitudinal null geodesics, the geometry at
the limit is six-dimensional Minkowski spacetime. In the case of
radial null geodesics, $u=\pm r=x^-$ and so the Penrose limit metric
can be given explicitly. In particular we find
\begin{multline}
  ds^2= 2 dx^+ dx^-+\big[\big(\frac5{16}(H_1 H_2)^{-2}
    (\partial_-(H_1 H_2) )^2-\frac14(H_1 H_2)^{-1} \partial^2_-(H_1
    H_2)\big)  x^2 \\
    -\frac3{16}(H_1 H_2)^{-2} (\partial_-(H_1 H_2) )^2+ \frac14 (H_1
    H_2)^{-1} \partial^2_-(H_1 H_2)\big)   y^2 \big](dx^-)^2 +
    ds^2(\EE^4)~.
\end{multline}

The near horizon geometry of this string solution is $\AdS_3\times
S^3$. Thus the Penrose limit in this case is (i) six-dimensional
Minkowski spacetime if it is taken along a radial null geodesic or
(ii) $\CW_6$ if it is taken along a generic null geodesic. This is
the solution obtained in \cite{Meessen}.

\section{Cosmological Penrose limits}
\label{sec:frw}

So far we have considered Penrose limits of space-times corresponding
to (intersecting) branes and their near horizon limits. But of course
the Penrose limit construction is more general than that. We will now
consider Penrose limits of cosmological Friedmann-Robertson-Walker
(FRW) space-times.  For definiteness we consider four-dimensional
models but qualitatively nothing changes in other dimensions.

The FRW metric is
\begin{equation}
 \label{eq:frwmet}
-dt^{2} + a(t)^2 d\tilde{s}^2~,
\end{equation}
where $d\tilde{s}^2$ is the line element of a maximally symmetric
space which can be written as
\begin{equation}
d\tilde{s}^2=dr^{2} + f_{k}(r)^{2}d\Omega_{2}^{2}~,
\end{equation}
where $f_{k}(r) = r, \sin r, \sinh r$ for $k=0,\pm 1$ respectively.

A FRW cosmological model is determined by specifying $k$, the
cosmological constant $\Lambda$, and the perfect fluid matter content,
usually characterised by the equation of state $p=w\rho$ relating the
energy-density $\rho$ and the pressure-density $p$ of the perfect
fluid, with $0\leq w \leq 1$ for not too exotic matter.

The cosmic scale factor $a(t)$ is then determined by solving the
Friedmann equations.  Typically these equations cannot be solved in
closed form when both $\Lambda$ and $\rho$ are non-zero. We will
therefore set $\Lambda = 0$ (for $\Lambda \neq 0$ but $\rho=0$ we
obtain (A)dS space-times whose Penrose limits we already know).

For $k=0$ and any $w$, the solution is
\begin{equation}
k=0 \Rightarrow a(t) = \beta t^{\alpha}~,
\end{equation}
where
\begin{equation}
\alpha = \tfrac23 (1+w)^{-1}
\end{equation}
and $\beta$ is a constant.

Without loss of generality (covariance of the Penrose limit and
isotropy of space), one can choose the null geodesics to lie in the
$(t,r)$-plane.
These null geodesics are characterised by the equations
(an overdot now denotes differentiation with respect to the affine
parameter $\tau$ which will become the coordinate $u$)
\begin{align}
 \label{eq:frwgeo}
\dot{t}^{2} &= a^{2} \dot{r}^{2}\\
\dot{r} &= c a^{-2}~,
\end{align}
where $c$ is a constant which without loss of generality can be set
equal to $c=1$ by rescaling the affine parameter. This leads to
\begin{equation}
\dot{t} = a^{-1}~.
\end{equation}
With $a(t)$ as above, this integrates to
\begin{align}
t(\tau) &= T \tau^{\frac{1}{\alpha +1}}\\
r(\tau) &= R \tau^{\frac{1-\alpha}{1+\alpha}}\\
a(\tau) &= A \tau^{\frac{\alpha}{\alpha +1}}~,
\end{align}
where $T,R,A$ are constants satisfying a quadratic relationship
following from \eqref{eq:frwgeo} whose precise numerical values are
irrelevant for the following.

Having obtained this congruence of null geodesics we now change coordinates
$(t,r)\to (u,v)$ such that
\begin{equation}
 \begin{split}
\d_{u} &= \dot{t}\d_{t} + \dot{r}\d_{r}\\
         &= a(t)^{-1}\d_{t} + a(t)^{-2}\d_{r}~,
 \end{split}
\end{equation}
and such that $g_{uv}=1$ and $[\d_u,\d_v]=0$. A possible choice is
\begin{equation}
\d_{v} = \d_{r}~.
\end{equation}
This integrates to
\begin{align}
t(u,v)&= \int \frac{du'}{a(u')}\\
r(u,v)&= \int \frac{du'}{a(u')^{2}} + v~,
\end{align}
the FRW metric \eqref{eq:frwmet} takes the desired form
\begin{equation}
ds^{2} = 2 dudv + a(u)^{2} (dv)^{2} +  a(u)^{2} r(u,v)^{2} d\Omega_{2}^{2}~,
\end{equation}
and the Penrose limit is (in Rosen coordinates)
\begin{equation}
ds^{2} =  2 dudv + a(u)^{2} r(u,0)^{2} ds^{2}(\EE^2) ~.
\end{equation}
Performing the integral defining $r(u)$ one finds that, up to an irrelevant
scaling of the $\tilde{x}$-coordinates of $\EE^2$,
\begin{equation}
ds^{2} = 2 dudv + u^{\frac{2}{1+\alpha}}ds^2(\EE^2)~.
\end{equation}
To pass to Brinkman coordinates, one sets
\begin{align}
u &= x^{-}\\
v &= x^{+} + \tfrac{1}{2(1+\alpha)}(x^-)^{-1} x^{2}\\
\tilde{x}^{i} &= (x^-)^{-\frac{1}{1+\alpha}} x^{i} ~,
\end{align}
and then finds
\begin{equation}
ds^{2} = 2 dx^- dx^+ + A(x^-,x)(dx^-)^{2} + ds^2(\EE^2)~,
\end{equation}
where
\begin{equation}
 A(x^-,x)= - [ \tfrac{1}{1+\alpha} - \tfrac{1}{(1+\alpha)^{2}} ]
 (x^-)^{-2} x^{2}~.
\end{equation}

We note that $A(x^-,x)$ has the characteristic $(x^-)^{-2}$-dependence
also encountered in the Penrose limit of the fundamental string
\eqref{eq:PLfsA}.
For all `physical' values of $\alpha$, $1/3 \leq \alpha \leq 2/3$,
$A(x^-,x)$ is negative definite.  The $\alpha$-dependence (i.e., the
dependence on the equation of state parameter $w$) of this metric
cannot be scaled away by scaling $(x^-,x^+)$ because precisely for
$A(x^-,x)\sim (x^-)^{-2}$, the combination $A(x^-,x) (dx^-)^2$ is
scale invariant.

According to \cite[\S\S 10.2,21.5]{ExactSolutions}, a pp-wave with this
profile is generically of type $G_{6}$ and $V_{4}$.  This means that there
is a six-dimensional group of isometries with four-dimensional
orbits, i.e., with the isometry group acting transitively on space-time,
in contrast to the FRW case which is of cohomogeneity one.  Thus the
limiting space-time is, while not Lorentzian symmetric, at least
Lorentzian homogeneous.

Actually this particular metric has a seven-dimensional isometry group
generated by the six Killing vectors inherited from the FRW
space-time, and the additional Killing vector
\begin{equation}
x^+\d_+ - x^-\d_-
\end{equation}
generating the isometry $(x^+,x^-)\to (cx^+,c^{-1}x^-)$, i.e., an
infinitesimal boost.  From the point of view of a general pp-wave
spacetime, these seven Killing vectors arise as the $2D{-}3{=}5$
Heisenberg algebra Killing vectors \eqref{eq:heisenberg} of a generic
pp-wave, plus the above scale invariance, plus the `accidental'
rotation symmetry in the two transverse dimensions.

\section{The Penrose limit and isometric embeddings}
\label{sec:embeddings}

In this section we will make some preliminary remarks about Penrose
limits and isometric embeddings.  We will show, in the context of a
toy model, that the Penrose limit of a spacetime is induced by a
generalised Penrose limit in a space with two times.  This hints at
the existence of a generalised Penrose limit for pseudoriemannian
spaces with arbitrary signature.  In the process we show how to embed
any Cahen--Wallach space with signature $(1,D-1)$ isometrically as the
intersection of two quadrics in a flat space with signature $(2,D)$.

\subsection{A toy model}

We start with a ``toy model'' corresponding to the near horizon
geometry of the Reissner--Nordström black hole in four-dimensional
$N{=}2$ supergravity, namely $\AdS_2 \times S^2$ with equal radii of
curvature.  As discussed in Section~\ref{sec:NHPL}, we can embed
$\AdS_2 \times S^2$ as the intersection of two quadrics in
$\EE^{2,4}$.  Indeed, this case corresponds to $p=0$, $D=4$ (and hence
$n=2$) in the notation of Section~\ref{sec:metrics}.  An explicit
parametrisation is given by
\begin{equation}
  \begin{split}
    X^0 = R\cos\tau~,\quad X^1=R \sinh\beta \sin\tau
    \quad\text{and}\quad X^2 = R \cosh\beta \sin\tau\\
    X^3 = R\cos\psi~,\quad X^4=R \sin\psi\cos\theta
    \quad\text{and}\quad X^5 = R \sin\psi \sin\theta~,
  \end{split}
\end{equation}
where $R$ is the common radius of curvature of the two spaces.  The
Penrose limit starts by rescaling the coordinates as follows
\begin{equation}
  \psi = \half (u + \Omega^2 v)~,\quad \tau = \half( u - \Omega^2 v)~,
  \quad \theta = \Omega y^1 \quad\text{and}\quad \beta = \Omega y^2~,
\end{equation}
in terms of which the ambient coordinates develop a dependence on
the scaling parameter $\Omega$, as follows
\begin{equation}
  \begin{split}
    X^0(\Omega) &= R \cos \half (u-\Omega^2 v)\\
    X^1(\Omega) &= R \sinh(\Omega y^2) \sin \half (u-\Omega^2 v)\\
    X^2(\Omega) &= R \cosh(\Omega y^2) \sin \half (u-\Omega^2 v)\\
    X^3(\Omega) &= R \cos \half (u+\Omega^2 v)\\
    X^4(\Omega) &= R \cos(\Omega y^1) \sin \half (u+\Omega^2 v)\\
    X^5(\Omega) &= R \sin(\Omega y^1) \sin \half (u+\Omega^2 v)~.
  \end{split}
\end{equation}
This suggests the following rescaling of the ambient coordinates
\begin{equation}
  \begin{aligned}[m]
    X^3 + X^0 &= U^1\\
    X^3 - X^0 &= \Omega^2 V^1\\
    X^1 &= \Omega Y^1
  \end{aligned}
  \qquad\qquad
  \begin{aligned}[m]
    X^4 + X^2 &= U^2\\
    X^4 - X^2 &= \Omega^2 V^2\\
    X^5 &= \Omega Y^2
  \end{aligned}~,
\end{equation}
which induces a homothety of the flat metric in $\EE^{2,4}$.  The new
coordinates $\{U^\mu,V^\mu,Y^i\}$ depend on $\Omega$.  To leading
order in $\Omega$ the embedding becomes
\begin{equation}
  \begin{aligned}[m]
  U^1 &= 2 R \cos \half u \\
  V^1 &= -R v  \sin \half u \\
  Y^1 &= R y^1 \sin \half u
  \end{aligned}
  \qquad\qquad
  \begin{aligned}[m]
  U^2 &= 2 R \sin \half u \\
  V^2 &= R \left( v \cos \half u - \half |y|^2 \sin \half u \right) \\
  Y^2 &= R y^2 \sin \half u~,
  \end{aligned}
\end{equation}
where $|y|^2 = (y^1)^2 + (y^2)^2$ and where all the omitted terms go
to zero as $\Omega \to 0$.  In the limit we obtain an embedded
submanifold of $\EE^{2,4}$ with ambient coordinates
$\{U^\mu,V^\mu,Y^i\}$ and metric
\begin{equation}
  G = dU^1 dV^1 + dU^2 dV^2 + (dY^1)^2 + (dY^2)^2~,
\end{equation}
corresponding to the intersection of two quadrics
\begin{equation}
  (U^1)^2 + (U^2)^2 = 4 R^2 \quad\text{and}\quad (Y^1)^2 + (Y^2)^2 +
  U^1 V^1 + U^2 V^2 = 0~.
\end{equation}
The induced metric is none other but a Cahen--Wallach Hpp-wave (in
Rosen coordinates)
\begin{equation}
  R^{-2} \bar g = du dv + (\sin \half u)^2 \left( (dy^1)^2 + (dy^2)^2
  \right)~.
\end{equation}
In terms of Brinkman coordinates, the embedding takes the simpler form
\begin{equation}
  \begin{aligned}[t]
  U^1 &= 2 R \cos x^- \\
  V^1 &= - R x^+ \sin x^- - \half R |x|^2 \cos x^- \\
  Y^1 &= R x^1
  \end{aligned}
  \qquad
  \begin{aligned}[t]
    U^2 &= 2 R \sin x^- \\
    V^2 &= R x^+ \cos x^- - \half R |x|^2 \sin x^- \\
    Y^2 &= R x^2~,
  \end{aligned}
\end{equation}
where $|x|^2 = (x^1)^2 + (x^2)^2$ and the induced metric is
\begin{equation}
  R^{-2} \bar g = 2 dx^+ dx^- - |x|^2 (dx^-)^2 + (dx^1)^2 + (dx^2)^2~.
\end{equation}

\subsection{Isometric embeddings of Cahen--Wallach spaces}

This toy model teaches us how to embed a general $D$-dimensional
Cahen--Wallach metric (Hpp-wave with constant $A_{ij}$)
isometrically\footnote{The existence of this embedding is stated in
  \cite{CahenWallach}, but the suggested embedding in that paper does
  not seem to work.  The isometric embedding of pp-waves is itself not
  new.  It has been discussed in the four-dimensional context in
  \cite{RosenEmbeddings,CollinsonEmbeddings} and is reviewed in
  \cite[\S 32]{ExactSolutions}.} in $\EE^{2,D}$.  Indeed, consider the
metric
\begin{equation}
  \label{eq:HppWaveMetric}
  g = 2 dx^+ dx^- + A(x) (dx^-)^2 + \sum_i dx^i dx^i~,
\end{equation}
where $A(x) = \sum_{i,j} A_{ij} x^i x^j$, where $A_{ij}$ is a constant
symmetric matrix, not necessarily positive-definite or even
non-degenerate.  Let us introduce coordinates $U^\mu,V^\mu,Y^i$ for
$\mu=1,2$ and $i=1,2,\dots,D-2$ for $\EE^{2,D}$ in terms of which the
flat metric is written as
\begin{equation}
  \label{eq:flatmetric}
  G = \sum_{\mu=1}^2 dU^\mu dV^\mu + \sum_{i=1}^{D-2} (dY^i)^2~.
\end{equation}
Consider the codimension-two submanifold of $\EE^{2,D}$ cut out by the
intersection of the two quadrics
\begin{equation}
 \sum_{\mu=1}^2 (U^\mu)^2 = 4 \qquad\text{and}\qquad
  A(Y) - \sum_{\mu=1}^2 U^\mu V^\mu = 0~,
\end{equation}
where $A(Y) = \sum_{i,j} A_{ij} Y^i Y^j$.  We can parametrise this
submanifold in terms of coordinates $x^\pm, x^i$ for
$i=1,2,\dots,D-2$ as follows
\begin{equation}
  \begin{split}
    U^1 = 2 \cos x^- \qquad U^2 = 2 \sin x^- \qquad Y^i = x^i\\
    V^1 = - x^+ \sin x^- + \half A(x) \cos x^- \qquad V^2 = x^+ \cos
    x^- + \half A(x) \sin x^-~,
  \end{split}
\end{equation}
and one sees that the induced metric coincides with
\eqref{eq:HppWaveMetric}.

\subsection{A generalised Penrose limit}

The above toy model teaches us another thing, namely it suggests the
existence of a generalisation of the Penrose limit in spaces with two
times.  (In fact, it should clear from our discussion below that this
admits a straight-forward generalisation to spaces with more than two
times.)  Consider the following diagram illustrating the relationships
between $\AdS_2 \times S^2$, its Penrose limit denoted by $\CW_4$ and
the embedding space
\begin{equation}
  \begin{CD}
    @. \\
    \EE^{2,4} @.  \EE^{2,4}\\
    @AAA          @AAA \\
    \AdS_2 \times S^2 @>\text{Penrose limit}>> \CW_4\\
    @.
  \end{CD}
\end{equation}
where the vertical arrows are the isometric embeddings described
above.  A natural question is whether there exists some limiting
procedure in $\EE^{2,4}$ which induces the Penrose limit when
restricted to $\AdS_2 \times S^2$.  In other words,
\begin{center}
  \emph{Is there an arrow $\EE^{2,4} \xrightarrow{?} \EE^{2,4}$ which
    completes the above diagram?}
\end{center}

{}From the discussion above it seems clear that this question has a
positive answer.  This limiting procedure involves, not a null
geodesic as in lorentzian signature, but rather a maximally isotropic
totally geodesic submanifold---in this case, a maximally isotropic
plane.\footnote{An isotropic submanifold is one on which the
  restriction of the metric is identically zero, and a maximally
  isotropic submanifold is an isotropic submanifold of maximal
  dimension.  In signature $(p,q)$ a maximally isotropic submanifold
  has dimension $\min\{p,q\}$, whence in $(2,D)$ signature it has
  dimension 2 (for $D\geq 2$)..  We will focus on this case, as it is
  the relevant case for our discussion.}

We can choose coordinates $U^\mu,V^\mu,Y^i$ for $\mu=1,2$ and
$i=1,2,\dots,D-2$ for $\EE^{2,D}$ in a neighbourhood of a totally
geodesic, maximally isotropic submanifold in such a way that the flat
metric of $\EE^{2,D}$ takes the form
\begin{equation}
  G = \sum_\mu dV^\mu \left(dU^\mu + \sum_\nu \alpha_{\mu \nu} dV^\nu
  + \sum_i \beta_{\mu i} dY^i\right) + \sum_{i,j} C_{ij} dY^i dY^j~,
\end{equation}
where $\alpha$, $\beta$ and $C$ can depend in principle on all the
coordinates.  We also demand that $C$ be positive-definite; although
this may limit the domain of validity of this coordinate system as
with the original Penrose limit.  The coordinates $U^\mu$ parametrise
a family of maximally isotropic surfaces labelled by $V^\mu$ and
$Y^i$.  The vectors $\d/\d U^\mu$ are null, orthogonal and
self-parallel, hence geodetic.  We can now rescale the coordinates in
the following way:
\begin{equation}
  U^\mu = u^\mu~,\quad V^\mu = \Omega^2 v^\mu \quad\text{and}\quad Y^i =
  \Omega y^i~.
\end{equation}
Substituting this into the metric $G$ yields a metric which depends on
$\Omega$ and such that the limiting metric
\begin{equation}
  \overline G = \lim_{\Omega\to 0} \Omega^{-2} G(\Omega)
\end{equation}
is well defined.  Taking the limit we find
\begin{equation}
  \overline G = \sum_\mu du^\mu dv^\mu  + \sum_{i,j} \overline
  C_{ij}(u) dy^i dy^j~.
\end{equation}

Now suppose that $M$ is an isometrically embedded submanifold of
$\EE^{2,D}$ with signature $(1,D-1)$.  A maximally isotropic plane
will cut $M$ generically in a null curve $\gamma$, if they meet at
all.  One might be surprised at this fact, since a two-plane and a
codimension-two submanifold will meet generically at a point, if they
meet at all; but in this case, since the surface is maximally
isotropic and the manifold has lorentzian signature, they must have a
direction in common at any point in their intersection.  If the curve
$\gamma$ is a null geodesic, then the generalised Penrose limit of
$\EE^{2,D}$ induces the Penrose limit of $M$ along $\gamma$.

We saw this above for the case of $\AdS_2 \times S^2$ embedded in
$\EE^{2,4}$.  In this case, the maximally isotropic surface was an
affine plane.  The same procedure works also for the Penrose limit of
$\AdS_3 \times S^3$, which is the near horizon geometry of the
self-dual string in $D=6$ $(2,0)$ supergravity, and for $\AdS_5\times
S^5$, which is the near horizon geometry of the D3 brane in IIB
supergravity.  It is conceivable that this also works in more
generality and if so, we will discuss this elsewhere.

A special case of this generalised Penrose limit has appeared in
\cite{ORS} in the context of exact string backgrounds given by WZW
models on nonsemisimple Lie groups admitting a bi-invariant metric
(see, e.g., \cite{FS3}).  As shown in \cite{ORS}, the Lie algebras of
some of these Lie groups can be obtained by a contraction of
semisimple Lie algebras.  This contraction is a special case of the
generalised Penrose limit alluded to above.  Indeed, let $G$ be a
compact simple Lie group and $H \subset G$ a subgroup.  The
Cartan--Killing form defines on $G$ a bi-invariant metric $g$, which
restricts nondegenerately to a bi-invariant metric $h = g\bigr|_H$ on
$H$.  Now consider the product group $G \times H$ with the product
metric $g\oplus - h$, where we have changed the sign of the metric in
the second factor.  We remark that this metric is again bi-invariant.
Consider now the submanifold $H \subset H \times H \subset G \times
H$ given by the diagonal embedding.  Since it is a Lie subgroup, it is
totally geodesic (in fact, relative to any bi-invariant metric), and
by virtue of our choice of metric it is maximally isotropic.  If we
now perform the generalised Penrose limit of $G\times H$ along
$H$ we obtain a non-semisimple Lie group with a bi-invariant metric.
This is essentially the construction in \cite{ORS}, which, as shown in
\cite{FSsug}, is an example of a more general construction of Lie
groups with bi-invariant metrics first discussed in
\cite{MedinaRevoy}.

\section{Worldvolume dynamics and Penrose limits}
\label{sec:wd}

The effect of the Penrose limit on the dynamics of brane probes with
worldvolume and Wess-Zumino couplings, like the fundamental string
\cite{GreenSchwarzString} and the M2-brane \cite{BSTSupermembrane}, has
been investigated in \cite{ShortLimits}. It was found that the Penrose
limit is a large tension limit for the probe.  Here we shall extend
this analysis for branes with fields of Born-Infeld type, such as
D-branes and the M5-brane.

The D$p$-brane tension in terms string units as a function of $p$ is
$T_p=\frac1{k_p(\alpha')^{\frac{p+1}{2}}}$ for some constant $k_p$
which depends of the string coupling constant.  The bosonic part of a
typical action of a D$p$-brane in a curved background
\cite{CvGNSW,APSDbrane1,APSDbrane2,BergTownDBranes} is
\begin{equation}
  I_p[g, B, C]=T_p\left(\int d^{p+1}\sigma e^{-\Phi} \sqrt{ g+ \eF} + \int
  e^{\eF}\wedge C\right)~,
\end{equation}
with $\eF=\alpha' F+ B$ and $C=\sum_k C_k$, where $F$ is the
Born--Infeld two-form field strength, $B$ is the NSNS
two-form gauge potential and $C_k$ are the RR k-form gauge
potentials. Next we set $\alpha'\sto \Omega^2\alpha'$.  Assuming that
the Born--Infeld field $F$ scales as $F\sto F$, the Dp-brane action
can be written as
\begin{multline}
  I_p[\Omega^{-2} g, \Omega^{-2}B,\Omega^{-k} C] = \frac1{k_p
    (\alpha')^{\frac{p+1}2}} \biggl( \int d^{p+1}\sigma e^{-\Phi}
  \sqrt{\Omega^{-2} (g+B)+\alpha'
    F} \\
  + \int \biggl[ \sum_k e^{\Omega^{-2} B +\alpha' F}\wedge \Omega^{-k}
  C_k\biggr]_{p+1}\biggr)~.
\end{multline}
Adapting coordinates for the Penrose limit and taking $\Omega\ll 1$, the
D$p$-brane action can be expanded as
\begin{equation}
  I_p[\Omega^{-2} g, \Omega^{-2}B,\Omega^{-k} C]=I_p[\bar g, \bar B,
  \bar C]+O(\Omega)~,
\end{equation}
where $\bar g, \bar B, \bar C$ are the fields at the Penrose limit.
Therefore we conclude that Dp-branes at the large tension limit
propagate in the Penrose limit of the associated spacetime. Since
there are different Penrose limits depending on the choice of the null
geodesic, there are different ways of taking the large tension limit
of a D$p$-brane.

Let us elaborate on this. We consider two null geodesics $\gamma^k$,
$k=1,2$ with distinct Penrose limits. From the starting metric $g$ we
pass to the two distinct adapted metrics $g^k$ (and likewise for the
fields $B$ and $C$) and from there to the two one-parameter families of
metrics $g^k_\Omega$. For all $\Omega >0$, $g^1_\Omega$ and $g^2_\Omega$
are isometric, being related by a coordinate transformation which will in
general depend in a complicated way on $\Omega$. Nevertheless, because the
brane probe action is generally covariant (and gauge invariant), we have
\begin{equation}
I_p[g^1_\Omega,B^1_\Omega,C^1_\Omega] =
I_p[g^2_\Omega,B^2_\Omega,C^2_\Omega]
\end{equation}
for all $\Omega >0$. However, this coordinate transformation becomes
singular as $\Omega\to 0$ and the $\Omega$- (or large tension) expansion
is around two distinct backgrounds. Moreover, even though, as we have
just seen, the two actions are non-perturbatively (i.e., for finite
$\Omega$) equivalent, the higher order terms in $\Omega$ will not agree
order by order in the perturbative expansion because the two actions
are related by an $\Omega$-dependent coordinate transformation. Hence
what is a perturbative effect in one perturbation expansion, may be
non-perturbative in another.

Combining the above result with that of \cite{ShortLimits}, we
conclude that a string theory in a supergravity background at large
tension as taken by a Penrose limit is equivalent to string theory at
a Penrose limit of that background. Since it is simpler to investigate
string theory in pp-wave type of spacetimes, it is expected that
string theory can be solved exactly at some Penrose limits. For the
maximally supersymmetric case of \cite{ShortLimits}, this has already
been done \cite{MetsaevIIB} and some of the consequences have
subsequently been explored in \cite{MaldaPL}.

Next we turn to investigate the Penrose limit of a M5-brane probe in
an eleven-dimensional spacetime \cite{HoweSezginBranes1,
  HoweSezginBranes2, HSW, BLNPST, APPS5Brane}. To describe the field
equations of the M5-brane in a general supergravity background as
given in \cite{HSW}, we begin with some definitions. We denote with
$\gamma_{\mu\nu}=e_\mu^a e_\nu^b n_{ab}$ the induced metric on the
worldvolume. Next $\eE^{\underline a}_\mu= \partial_\mu X^M
E_M^{\underline a}$, where $X$ is the embedding of map of the M5-brane
into spacetime and $E_M^{\underline a}$ is the spacetime frame. Let
$\eh$ be the self-dual three form and $\eC=\kappa_2 F_3+C_3$, where
$d\eC=-\frac14 F_4$, i.e., $F_3$ is the worldvolume field on the
M5-brane, $\kappa_2$ is the inverse of the M2-brane tension and $C_3$
is the eleven-dimensional supergravity three-form gauge potential.  We
also have
\begin{equation}
  \begin{aligned}
    \eC_{abc}&= m_a{}^d m_b{}^e \eh_{cde}\\
    m_a{}^b&=\delta_a{}^b-2\eh_{acd}\eh^{bcd}
  \end{aligned}
\end{equation}
and
\begin{equation}
  G^{\mu\nu}= (m^2)^{ab} e_a^\mu e_b^\nu\ .
\end{equation}
The field equations of the M5-brane are
\begin{equation}
  \begin{aligned}
    G^{\mu\nu} \nabla_\mu {\eC}_{\nu\rho\sigma}&=0
    \\
    G^{\mu\nu} \nabla_\mu {\eE}^{\underline a}&=(1-\frac23 \rm{tr} k^2)
    \epsilon^{\nu_1\dots \nu_6}\\
    & (\tfrac1{720} {}^*F^{\underline a}{}_{\nu_1\dots \nu_6}
    +\frac23 F^{\underline a}{}_{\nu_1\dots \nu_3} {\eC}_{\nu_4\nu_5\nu_6})
    (\delta^{\underline a}{}_{\underline c}-
    {\eE}^\mu_{\underline c}{\eE}_\mu^{\underline a})
  \end{aligned}
\end{equation}
where $\nabla$ is the covariant derivative with respect to the
induced metric and $k_a{}^b=\eh_{acd} \eh^{bcd}$.

To take the Penrose limit, we adopt the appropriate light cone
coordinates and scale the spacetime fields as usual $g\sto \Omega^{-2}
g$ and $C_3\sto \Omega^{-3} C_3$. Consequently, $E^{\underline a}\sto
\Omega^{-1}E^{\underline a}$, the induced metric scales as $\gamma\sto
\Omega^{-2} \gamma$ and $e^a\sto \Omega^{-1} e^a$.  It is then easy to
see that the metric $G$ scales as $G\sto \Omega^{-2} G$.  Provided that
$\kappa_2\sto \Omega^3 \kappa_2$ and $F_3\sto F_3$, the field equations
for the embedding scalars scales with weight $1$ while the field
equation for the two-forms gauge potential scales with weight $-1$.
Therefore a solution of the M5-brane field equations remains a
solution in the Penrose limit.

Combining the above result with that of \cite{ShortLimits}, we have
found that the Penrose limit is a large membrane tension limit in
M-theory. Placing the M-branes in an eleven-dimensional background and
taking the large M2-brane tension limit and so that of the
M5-brane\footnote{The tension of the M5-brane is the square of that of
  the M2-brane.}, the dynamics of the M-branes is that of M-branes in
a Penrose limit of the original background. Again there are different
ways of taking the large M2-brane tension limit depending on the
choice of null geodesic associated with the Penrose limit.

\section*{Acknowledgments}

It is a pleasure to thank Chris Hull for his collaboration at the
initial stages of this work. We also thank Gary Gibbons and Paul
Townsend for helpful comments, and David Calderbank for sharing his
geometric expertise and for bringing \cite{KostantHol} to our
attention.  This work was initiated while two of us (MB,JMF) were
participating in the programme \emph{Mathematical Aspects of String
  Theory} at the Erwin Schrödinger Institute in Vienna.  We are
grateful to the ESI for their support.  The research of MB is
partially supported by EC contract CT-2000-00148.  JMF is a member of
EDGE, Research Training Network HPRN-CT-2000-00101, supported by The
European Human Potential Programme.  The research of JMF is partially
supported by the EPSRC grant GR/R62694/01.  In addition, JMF would
like to acknowledge a travel grant from PPARC.  GP is supported by a
University Research Fellowship from the Royal Society.  This work is
partially supported by SPG grant PPA/G/S/1998/00613.

\bibliographystyle{utphys}
\bibliography{AdS3,AdS,ESYM,Sugra,Geometry,CaliGeo}

\end{document}